\documentclass[superscriptaddress, aps,  floatfix, prb,twocolumn]{revtex4-2}
\usepackage{epstopdf}  

\pdfoutput=1 %

\usepackage{graphicx}   

\usepackage{dcolumn}

\usepackage{hyperref}
\usepackage{bm}                 
\usepackage{amsmath}
\usepackage{amsfonts}
\usepackage{revsymb}
\usepackage{color}

\def\wn/{\,cm$^{-1}$}
\def\area/{\,cm$^{-2}$}
\def\cubic/{$_\mathrm{c}$}
\def\DM/{Dzyaloshinskii-Moriya}

\def\bfo/{BiFeO$_3$}

\def\edc/{\ensuremath{\mathbf{E}}}
\def\eac/{\ensuremath{\mathbf{E}_\omega}}
\def\bdc/{\ensuremath{\mathbf{B}}}
\def\hdc/{\ensuremath{\mathbf{H}}}
\def\bac/{\ensuremath{\mathbf{B}_\omega}}
\def\hac/{\ensuremath{\mathbf{H}_\omega}}

\def\kvec/{\ensuremath{\mathbf{k}}}
\def\Pvec/{\ensuremath{\mathbf{P}}}
\def\Mvec/{\ensuremath{\mathbf{M}}}
\def\Tvec/{\ensuremath{\mathbf{T}}}
\def\Svec/{\ensuremath{\mathbf{S}}}

\newcommand{\e}[1]{\times 10^{#1}} 
\newcommand{\vect}[1]{\ensuremath{{\bf #1}}}

\def\ket#1{\left| #1 \right>}

\newcommand{\SpherHar}[2]{\ensuremath{Y_{#1}^{#2}}}
\newcommand{\SpherOp}[2]{\ensuremath{\mathcal{Y}_{#1}^{#2}}}

\def \vS {{\bf S}}
\def \vR {{\bf R}}
\def \ve {{\bf e}}
\def \vQ {{\bf Q}}
\def \vq {{\bf q}}

\def \ve {{\bf e}}
\def \vm {{\bf m}}

\def \mb {\mu_{\rm B}}

\def \zp {{\bf Z}}
\def \xx {{\bf x}}
\def \yy {{\bf y}}
\def \zz {{\bf z}}

\def \vM {{\bf M}}
\def \vP {{\bf P}}

\def \hi {{h_i}}

\def \XX {{\bf X}}
\def \YY {{\bf Y}}
\def \ZZ {{\bf Z}}
\def \mm {{\bf m}}

\graphicspath{{pdf/}}

\begin{document}


\title{The Magnetoelastic Distortion of Multiferroic \bfo/ in the Canted Antiferromagnetic State}
\author{T. R{\~o}{\~o}m}\thanks{toomas.room@kbfi.ee}
\author{J. Viirok}
\author{L. Peedu}
\author{U. Nagel}
\affiliation{National Institute of Chemical Physics and Biophysics, Akadeemia tee 23, 12618 Tallinn, Estonia}

\author{D. G.  Farkas}
\affiliation{Department of Physics, Budapest University of Technology and Economics and MTA-BME Lend\"ulet Magneto-optical Spectroscopy Research Group, 1111 Budapest, Hungary}
\affiliation{Condensed Matter Research Group of the Hungarian Academy of Sciences, 1111 Budapest, Hungary}

\author{D. Szaller}
\affiliation{Department of Physics, Budapest University of Technology and Economics and MTA-BME Lend\"ulet Magneto-optical Spectroscopy Research Group, 1111 Budapest, Hungary}
\affiliation{Institute of Solid State Physics, Vienna University of Technology, 1040 Vienna, Austria}

\author{V. Kocsis}
\affiliation{Department of Physics, Budapest University of Technology and Economics and MTA-BME Lend\"ulet Magneto-optical Spectroscopy Research Group, 1111 Budapest, Hungary}
\affiliation{RIKEN Center for Emergent Matter Science (CEMS), Wako 351-0198, Japan}

\author{S. Bord{\'a}cs}
\affiliation{Department of Physics, Budapest University of Technology and Economics and MTA-BME Lend\"ulet Magneto-optical Spectroscopy Research Group, 1111 Budapest, Hungary}
\affiliation{Hungarian Academy of Sciences, Premium Postdoctor Program, 1051 Budapest, Hungary}

\author{I. K{\'e}zsm{\'a}rki}
\affiliation{Department of Physics, Budapest University of Technology and Economics and MTA-BME Lend\"ulet Magneto-optical Spectroscopy Research Group, 1111 Budapest, Hungary}
\affiliation{Experimental Physics V, Center for Electronic Correlations and Magnetism, Institute of Physics, University of Augsburg, 86159 Augsburg, Germany}

\author{D. L. Kamenskyi}  
\author{H. Engelkamp}
\affiliation{High Field Magnet Laboratory (HFML-EMFL), Radboud University, Toernooiveld 7, 6525 ED Nijmegen, The Netherlands}

\author{M. Ozerov}
\author{D. Smirnov}
\author{J. Krzystek}
\affiliation{National High Magnetic Field Laboratory, Tallahassee, FL, 32310, USA}
\author{K. Thirunavukkuarasu}
\affiliation{Department of Physics, Florida A\&M University, FL, 32307, USA}

\author{Y. Ozaki}
\author{ Y. Tomioka}
\author{T. Ito}
\affiliation{National Institute of Advanced Industrial Science and Technology (AIST), Tsukuba, 305-8565 Ibaraki, Japan}

\author{T. Datta}
\affiliation{Department of Chemistry and Physics, Augusta University, 1120 15th Street, Augusta, Georgia 30912, USA}

\author{R. S. Fishman}\thanks{fishmanrs@ornl.gov}
\affiliation{Materials Science and Technology Division, Oak Ridge National Laboratory, Oak Ridge, Tennessee 37830, USA}

\date{\today}

\begin{abstract}
Using THz  spectroscopy, we show that the spin-wave spectrum of multiferroic \bfo/ in its high-field canted antiferromagnetic state is well described by a spin model that violates rhombohedral symmetry.
We demonstrate that the monoclinic distortion of the canted antiferromagnetic state is induced by the single-ion magnetoelastic coupling between the lattice and the two nearly anti-parallel spins.  
The revised spin model for \bfo/ contains two new single-ion anisotropy terms that violate rhombohedral symmetry and depend on the direction of the magnetic field.
\end{abstract}


\maketitle

\section{Introduction}

The room temperature multiferroic \bfo/ is one of the most technologically important materials in the rapidly expanding field of spintronics  \cite{Manipatruni2018,Manipatruni2019,Spaldin2019}, with applications to nanoelectronics \cite{Crassous2011,Yang2015} and photo-voltaics \cite{Yang09,Parsonnet2020}. 
One of the most useful ways to control the properties of \bfo/ thin films is through strain, which unwinds the cycloidal spin state and stabilizes a canted G-type antiferromagnet (AF) \cite{Bai2005,Ederer2005}.  Increasing epitaxial strain transforms the structure of thin films from rhombohedral to ``tetragonal-like'' monoclinic \cite{Bea2007,Iliev2010,MacDougall2012}.  Recent work on thin films \cite{Dixit2015,Chen2018} reveals that epitaxial strain can rotate the AF vector $\vS_1-\vS_2$ with respect to the electric polarization $\vP $.  Despite great interest in controlling its magnetic properties, comparatively little is known about the effects of magnetoelastic strain on bulk \bfo/ \cite{Kawachi2017}.

Magnetic properties of bulk materials are typically described by spin Hamiltonians with constant parameters.
Due to magnetostriction, however, those parameters may depend on  field and temperature.
In ferromagnetic (FM) materials, a large magnetic moment strains the crystal and the strain changes the spin couplings \cite{Callen1963,Callen1965,Callen1968,Alben1969}.  
Less is known about the effects of magnetostriction in AF materials or in  materials with weak FM moments, where the most notable manifestation of magnetostriction appears to be a spontaneous or field-induced spin reorientation \cite{Callen1968,Belov1987,Doerr2005}.

A small, canted magnetic moment less than 0.1\,$\mu_{\rm B}$ appears just above 18\,T
in the G-type AF phase of \bfo/ \cite{Kadomtseva2004,Tokunaga2010JPSJ,Park2011,Tokunaga2015a}.  
The spin model of \bfo/ in this canted phase is not well understood 
because high fields present challenges for both structural and spectroscopic probes. 
In this paper, we describe the THz absorption by spin waves in the high-field canted phase of \bfo/.  
Based on high-resolution measurements of the spin-wave frequencies, we show that the change of symmetry from rhombohedral to monoclinic activates two new coupling terms in the spin Hamiltonian.  
This work demonstrates that THz measurements can be used to determine the magnetoelastic coupling constants in the AF phase of a technologically-important material.

Following the appearance of the electric
polarization $\vP $ along one of the pseudo-cubic diagonals, 
the cubic symmetry of the perovskite structure of bulk \bfo/ is broken below $T_\mathrm{c} \approx 1100$ K \cite{Smith1968,Teague1970,Moreau1971}. A cycloidal spin state with wavevector $\vQ \perp \vP $ 
and spins predominantly in the plane defined by $\vQ $ and $\vP$ develops below
$T_\mathrm{N} \approx 640$ K \cite{Sosnowska1982,Lebeugle2008,Lee2008}.
In zero magnetic field, the cycloid has a wavelength of 62\,nm \cite{Sosnowska1982,Ramazanoglu2011,Herrero-Albillos2010,Sosnowska2011}.  
An applied magnetic field increases the wavelength and rotates $\vQ$ \cite{Bordacs2018}.
When a field applied perpendicular to $\vP$ exceeds $B_\mathrm{c}\approx18$\,T, the cycloid transforms into a G-type AF \cite{Ohoyama2011}.  Both the 
polarization $\vP$ and the magnetization $\vM$ exhibit step-like changes at $B_\mathrm{c}$ 
\cite{Popov1993,Kadomtseva2004}.

Although the crystal structure of bulk ferroelectric \bfo/ was first assigned to the rhombohedral space group $R3c$ \cite{Michel1969,Moreau1971,Kubel1990,Palewicz2007,Palewicz2010}, 
high-resolution structural studies suggested that the crystal symmetry might be monoclinic $Cc$ \cite{Sosnowska2012} or even lower, triclinic $P1$ \cite{Wang2013Structure}.
Additional evidence for broken symmetry comes from magnetostriction measurements: 
as the cycloid unwinds in a magnetic field, the contraction or expansion of the lattice depends on the direction of the applied field \cite{Kawachi2017}.
Based on a study of the THz absorption spectra
in the high-field canted AF phase,
this paper shows that magnetoelastic coupling transforms the crystal structure of \bfo/ from rhobomhedral to monoclinic.  A new microscopic model for the canted AF phase contains two new
single-ion anisotropy terms that break rhombohedral 
symmetry and depend on the orientation of the magnetic field.


Many years and tremendous effort have been spent constructing the spin model for bulk \bfo/.  
Much has been learned about the microscopic parameters by studying the spin-wave excitations of the cyloidal state using four different methods: inelastic neutron scattering (INS) \cite{Jeong2012,Matsuda2012,Xu2012,Jeong2014}, Raman \cite{Cazayous2008,Rovillain2010}, sub-millimeter wave electron spin resonance (ESR) \cite{Ruette2004}, and THz \cite{Talbayev2011,Nagel2013,Kezsmarki2015,Fishman2015} spectroscopies.
Because few sub-THz spectroscopic methods are compatible with high magnetic fields \cite{Ruette2004, Nagel2013}, much less is known about the spin-wave excitations in the canted AF state.

The two iron $S=5/2$ spins in the G-type AF structure produce two spin-wave modes, $\nu_1$ and $\nu_2$.
In earlier measurements,  
crystals were grown by the flux method, which provided platelets with a large surface parallel to $(001)$ crystal plane (pseudo-cubic notation).  
The lower frequency mode $\nu_1$ was then observed by ESR \cite{Ruette2004} and the upper mode $\nu_2$ by THz absorption spectroscopy \cite{Nagel2013} with field along $(001)$. 
The dependence of the mode frequencies on the field direction was not studied.

In the present study, a large single crystal grown using the floating zone method \cite{Ito2011} was cut into 0.5\,mm thick samples with large faces normal to \mbox{[1,-1,0]}, \mbox{[-1,-1,2]} and \mbox{[1,1,1]}.  
THz absorption measurements employed either Fourier transform far-infrared (FIR) or continuous wave (CW) spectroscopy.
FIR measurements were performed above 0.55\,THz  in a fixed magnetic field.
CW measurements were performed at a fixed frequency between 0.1 and 0.9\,THz by  sweeping the magnetic field, a method also called sub-millimeter wave ESR. 
Radiation propagated either parallel (Faraday configuration) or perpendicular (Voigt configuration) to the applied magnetic field. 
Descriptions of the experiment and measured spectra are provided in the Supplemental Material \cite{SM_BFO}.

\begin{figure}
	\includegraphics[width=0.5\columnwidth]{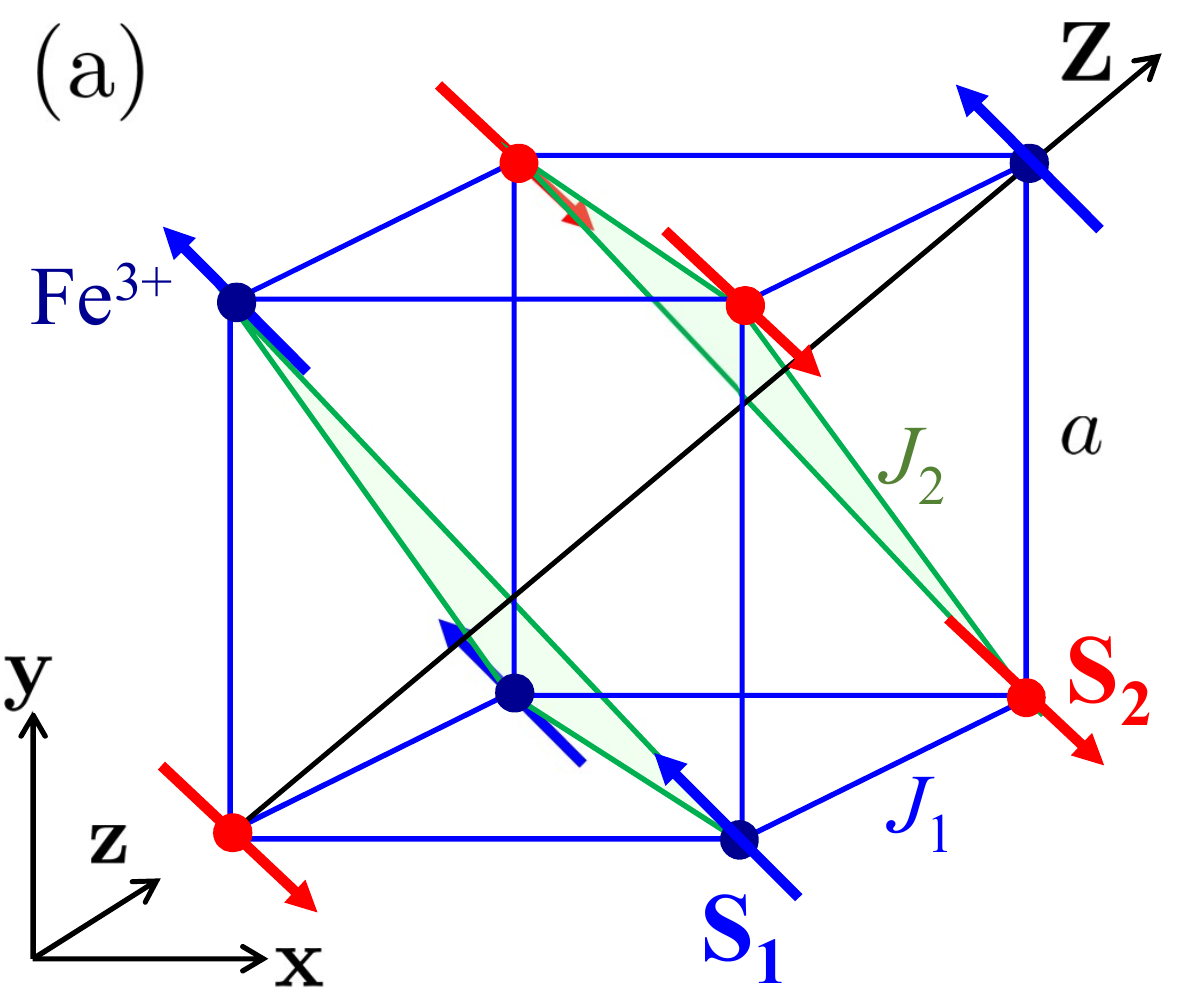}
	\includegraphics[width=0.4\columnwidth]{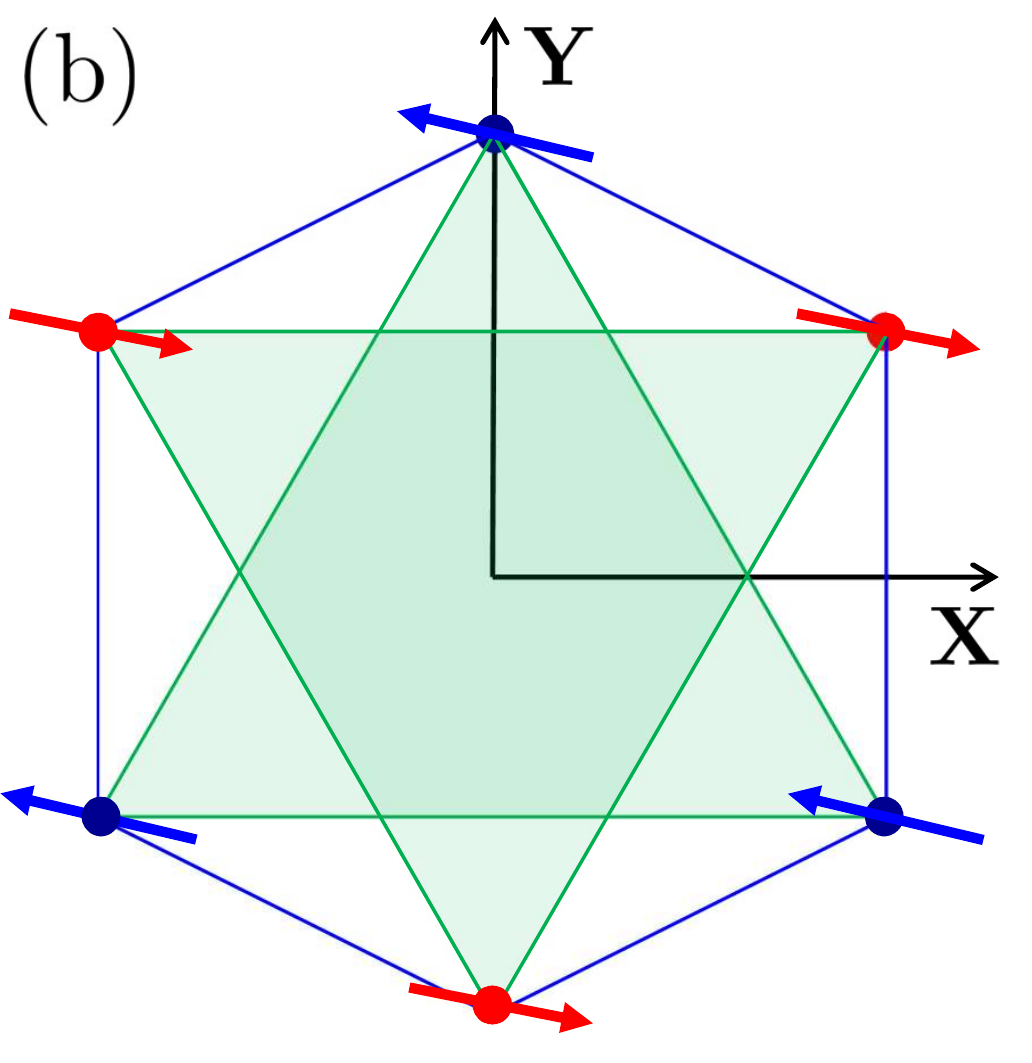}
	\caption{\label{fig:unit_cell}
		(Color online)  (a) The crystallographic pseudo-cubic unit cell of \bfo/.
		In the canted AF state, the magnetic unit cell is $2a\times 2a \times 2a$ with two spins, $\vS_1$ and $\vS_2$, per unit cell and pseudo-cubic
	 unit vectors $\xx = [1,0,0]$, $\yy = [0,1,0]$, and $\zz = [0,0,1]$. 
	(b) Fragments of two nearest-neighbor hexagonal planes viewed along $\ZZ$. 
	In the hexagonal planes (shaded triangles) normal to $\ZZ = [1,1,1]/\sqrt{3}$, the unit vectors are  
	 $\XX =[1,-1,0]/\sqrt{2}$ and  $\YY = [-1,-1,2]/\sqrt{6}$.
 }
\end{figure}

In low fields and temperatures, it is sufficient to treat the exchange, \DM/ (DM), and single-ion anisotropy parameters as  constants. 
At high magnetic fields, however, 
magnetoelastic coupling (magnetostriction) distorts the lattice and can change those parameters.
In a FM, Callen and Callen \cite{Callen1963,Callen1965,Callen1968,Alben1969} showed that
the driving force of magnetoelastic coupling is the macroscopic magnetic moment, which changes with temperature or in a magnetic field.
However, AFs do not have a net magnetic moment.
In \bfo/, the DM interaction and magnetic field cant the spins but the net magnetic moment is very weak.

To study the magnetoelastic properties of an AF, we expand the free energy in terms of the strain and the FM and AF ordering vectors, $\vS_1+\vS_2$ and $\vS_1-\vS_2$ \cite{Belov1987}. 
But calculating the spin-wave spectrum using this approach requires an exact spin-operator form for the magnetoelastic coupling.  This paper shows that THz spectroscopy can be used to narrow down the possible magnetoelastic coupling terms in the Hamiltonian and to determine the small coupling parameters.

This paper is divided into five sections.  
Section II describes the new spin model for \bfo/ in the high-field canted phase.  Predictions of that model are compared with THz measurements in Section III and the resulting model parameters are presented in Section IV.  
Section V contains a discussion and conclusion.  In the Supplemental Material \cite{SM_BFO}, we derive the possible magnetoelastic coupling terms consistent with monoclinic symmetry for \bfo/.

\section{Model}

We propose a new spin model for \bfo/  by applying the microscopic theory of Callen {\it et al.} \cite{Callen1963,Callen1968} to the canted AF state.
Even in the absence of a net magnetic moment, strain couples to the nearly collinear spins $\vect{S}_1$ and $\vect{S}_2$ in the magnetic unit cell.
Due to crystal fields, this magnetoelastic coupling affects the local single-ion anisotropy parameters of the spin Hamiltonian.  Details of this treatment are provided in the Supplemental Material \cite{SM_BFO}.

The earlier rhombohedral spin model of the cycloidal spin state contained two exchange constants, two DM  terms, and one anisotropy term:
\begin{eqnarray}
\label{Ham}
&&{\cal H}_{\mathrm{m}} = -J_1\sum_{\langle i,j\rangle }\vS_i\cdot \vS_j -J_2\sum_{\langle i,j \rangle'} \vS_i\cdot \vS_j  \\
&&+D_1\, \sum_{\langle i,j\rangle } (\zp \times {\bf e}_{i,j}/a) \cdot (\vS_i\times\vS_j) \nonumber \\
&& + D_2 \, \sum_{\langle i,j\rangle } \, (-1)^\hi \,\zp \cdot  (\vS_i\times\vS_j) -K_Z \sum_i  S_{iZ}^2 \nonumber \\
&& -\frac{1}{2}K_H \sum_i \left[(S_{iX}+i S_{iY})^6 + (S_{iX} - i S_{iY})^6\right] \nonumber \\
&&- g\mb B \sum_i \vm \cdot \vS_i\nonumber
\end{eqnarray}
where $\ve_{i,j} = a{\bf x}$, $a{\bf y}$, or $a{\bf z}$ connects the $S=5/2$ spin $\vS_i$ on site $\vR_i$ with the nearest-neighbor spin $\vS_j$ 
on site $\vR_j=\vR_i + \ve_{i,j}$.  The integer $h_i=\sqrt{3} \vR_i \cdot \zp /a$ is the hexagonal layer number.   
While the AF exchange $J_1$ couples nearest-neighbor spins along the edges of the cube, the AF exchange $J_2$ couples next-nearest neighbor spins along the cube face diagonals, Fig.\,\ref{fig:unit_cell}.  
Easy-axis anisotropy $K_Z$ lies along the polarization direction $\ZZ $.  
Hexagonal anisotropy \cite{Fishman2018PhysicaB, Fishman2018} $K_H$ pins the plane of the cycloid and the cycloidal wavevector $\vQ$ to one of the hexagonal axis $[1,-1,0]$, $[0,1,-1]$ or $[1,0,-1]$ perpendicular to $\ZZ $.  
The last term in (\ref{Ham}) is the interaction of spin $\vect{S}_i$ with a magnetic field $\vect{B} = B\vect{m}$.
We assume that the $g$ factor for the $S=5/2$ iron spins is isotropic with $g=2$.

Two DM interactions are produced by broken inversion symmetry.
While the first DM interaction $D_1$ determines the cycloidal period $\lambda $ \cite{Sosnowska1995},
the second DM interaction $D_2$ tilts the cycloid out of the plane defined by $\zp $ and the ordering wavevector $\vect{Q} \perp \vect{Z}$ \cite{Sosnowska1995,Kadomtseva2004,Pyatakov2009}.  
Because this tilt averages to zero over the length of the cycloid, \bfo/ has no spontaneous magnetic moment below $B_\mathrm{c}$.  
In the canted AF state above $B_\mathrm{c}$, \bfo/ has a small ferrimagnetic moment perpendicular to $\vP$ \cite{Kadomtseva2004,Tokunaga2010JPSJ,Park2011,Tokunaga2015a}.

Exchange parameters $J_1$ and $J_2$ are taken from INS \cite{Jeong2012, Matsuda2012, Xu2012}, which measured the spin-wave spectra over a wide range of energies and wavevectors.  
Because INS lacks sufficient wavevector resolution, the smaller DM and anisotropy terms were later estimated using THz absorption spectroscopy \cite{Nagel2013}.  
For convenience, Table \ref{tab:parameters} summarizes the values of these parameters and the experimental or theoretical methods used for their determination based on   
the properties of the cycloidal state assuming rhombohedral $R3c$ symmetry.

The spin model for \bfo/ undergoes significant simplifications in the high-field canted AF state.
Due to the steep dispersion $\omega = cq$ of photons, THz spectroscopy  measures the spin-wave frequencies at 
wavevector $q \ll 2\pi /a$.
With two spins in the cubic unit cell shown in Fig.1, $J_2$ does not contribute when $\vq \approx 0$  \footnote{Rhombohedral distortion of a cube, elongation along the body diagonal, introduces two different $J_2$ couplings. The difference between the two is not taken into account because the spin-wave frequencies do not depend on $J_2$ at $\vect{q}=0$}.
It is also easy to show that the first DM interaction $D_1$ has no effect on the mode frequencies in the canted AF state because it sums to zero.  
Taking $J_1\approx -5.3$ meV from INS measurements \cite{Jeong2012, Matsuda2012, Xu2012}, 
$\mathcal{H}_\mathrm{m}^{\mathrm{AF}}$ only depends on the DM interaction parameter $D_2$ and the anisotropy parameters $K_Z$ and $K_H$:
\begin{eqnarray}
\label{Ham_magn_AF}
&&{\mathcal H}_\mathrm{m}^{\mathrm{AF}}  = - J_1\sum_{\langle i,j\rangle }\vS_i\cdot \vS_j \\
&&+ D_2 \, \sum_{\langle i,j\rangle } \, (-1)^\hi \,\zp \cdot  (\vS_i\times\vS_j)\nonumber -K_Z \sum_i  S_{iZ}^2 \nonumber \\
&& -\frac{1}{2}K_H \sum_i \left[(S_{iX}+i S_{iY})^6 + (S_{iX} - i S_{iY})^6\right] \nonumber \\
&& - g\mb B \sum_i \vm \cdot \vS_i. \nonumber
\end{eqnarray}
Hexagonal anisotropy is the weakest interaction in this Hamiltonian with $K_HS^6 < 10^{-3}$ meV.
Since $\vert D_2 \vert \gg K_Z$, the spins lie primarily in the $XY$ plane with $\langle S_{iZ}\rangle \approx 0$.
The spin canting induced by the DM interaction and by magnetic fields \cite{Kawachi2017} up to about 35 T is less than $2^\circ$.
Consequently, the zero-order spin state is $\vect{S}_1\approx -\vect{S}_2$  and $(\vect{S}_1-\vect{S}_2) \perp \vect{B}$.

Because the spins are perpendicular to the field $\vect{B}$, the magnetoelastic strain depends on the field direction $\vect{m}$.
The equilibrium strain is solved by minimizing the elastic and magnetoelastic energies for field directions $\XX$ and $\YY$, see Supplemental Material \cite{SM_BFO}.
When $\vect{B}\parallel \ZZ$, there is no preferred orientation for the spins in the hexagonal plane and the strain vanishes.  For $\vect{B}\parallel \XX$, the zero-order spin state has $(\vect{S}_1-\vect{S}_2) \parallel \YY$.
For $\vect{B}\parallel \zz$ or $\parallel \YY$, the zero-order spin state has $(\vect{S}_1-\vect{S}_2) \parallel \XX$ because both $\zz $ and $\YY$ are perpendicular to $\XX =(\xx -\yy)/\sqrt{2}$.

Both the strain and the unit vectors $\XX_n$ and $\YY_n$
are determined by the field direction $\vect{m}$.  Hence, our analysis would be the same for $\vect{B}$ along any 
cubic axis.  If $\vect{B}\parallel \xx $, then the spins $\vS_1$ and $\vS_2$
would point (approximately) along $\pm \XX_3$ with $\XX_3 \equiv [0,1,-1]/\sqrt{2} \perp \xx $.  
If $\vect{B}\parallel \yy$, then the spins would point 
along $\pm \XX_2$ with $\XX_2 \equiv [-1,0,1]/\sqrt{2} \perp \yy $.  For specificity, we treat
the case $\vect{B}\parallel \zz $ with $\XX = \XX_1 \equiv [1,-1,0]/\sqrt{2}\perp \zz$.  In all cases,
$\YY_n = \ZZ \times \XX_n $.

The new spin state and spin-wave frequencies are modeled by the Hamiltonian
\begin{equation}\label{eq:ham_fit}
{\mathcal H}={\mathcal H}_\mathrm{m}^{\mathrm{AF}} + \sum_{i} \mathcal{H}_\mathrm{me}^i(\vect{m}),
\end{equation}
where the new strain-induced Hamiltonian for the $i$-th spin is
\begin{eqnarray}\label{eq:me_short}
	&&\mathcal{H}_\mathrm{me}^i(\vect{m}) = -K_{A,1}^{(\vect{m})} S_{iY} S_{iZ}	-K_{A,2}^{(\vect{m})} (S_{iX}^2-S_{iY}^2)\\
	&&-K_{E,1}^{(\vect{m})} S_{iY} S_{iZ}^3  -K_{E,2}^{(\vect{m})} (S_{iX}^2-S_{iY}^2)S_{iZ}^2\nonumber\\
	&&-K_{E,3}^{(\vect{m})} (S_{iX}^4 + S_{iY}^4 - 6 S_{iX}^2 S_{iY}^2),\nonumber
\end{eqnarray}
where the single-ion anisotropy constants depend on the field orientation $\vect{m}$.
As shown in the Supplemental Material, the two strains  $\epsilon_1^{\gamma,1}=\frac{1}{2}[\epsilon_{XX}- \epsilon_{YY} ]$ and $\epsilon_1^{\gamma,2}=\epsilon_{YZ}$ couple to the zero-order spin state \cite{SM_BFO}.

\section{Comparison with THz measurements}

For each field direction and magnitude, the energy $E = \langle {\mathcal H} \rangle $ was minimized as a function of angles $\theta_i$ and $\phi_i$ for the two spins $\vect{S}_i=S(\cos \phi_i \sin \theta_i \vect{X}+ \sin \phi_i \sin \theta_i \vect{Y}+\cos \theta_i  \vect{Z})$ in the unit cell.
Linear spin-wave theory was then used to evaluate the two spin-wave mode frequencies, which were compared with the measured frequencies.
This loop was repeated by varying the Hamiltonian parameters until a minimum $\chi^2$ was achieved \cite{SWBook2018}.

\begin{figure*}[!thbp]
	\includegraphics[width=1\textwidth]{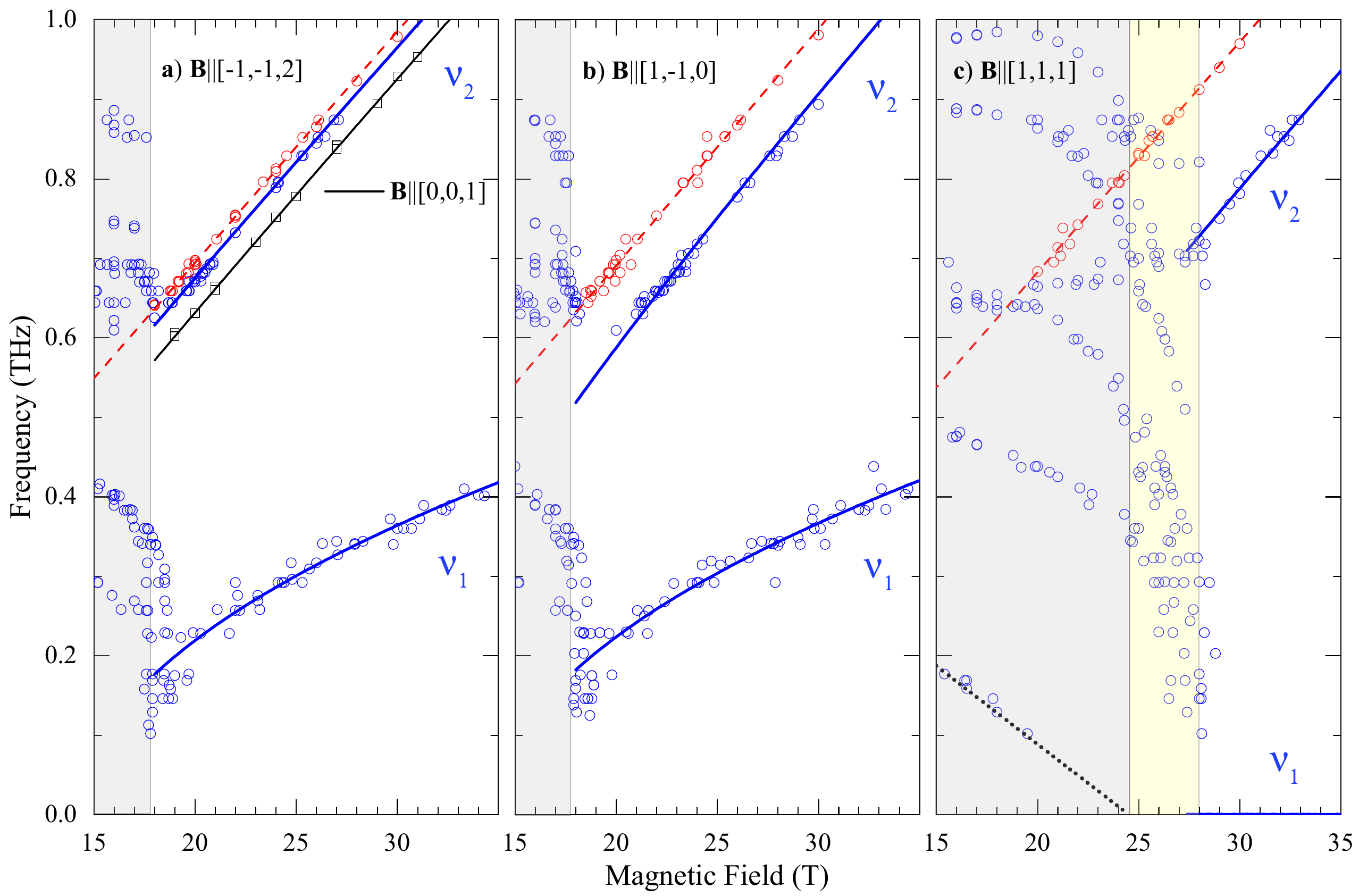}  
	\caption{\label{fig:Magnetic_spectra} Spin-wave frequencies at liquid He temperature for field along (a) $\YY$ (blue circles and blue lines) and $\zz$ (black squares and black line, data from \cite{Nagel2013}), (b) $\XX$, or  (c) $\ZZ$.
	Experimental points are marked by blue circles and black squares. 
	Solid lines are the best fits for spin-wave modes $\nu_1$ and $\nu_2$ using Eq.\,(\ref{eq:ham_fit}). 
	The dashed red line is the linear fit of impurity peak positions, red circles. 
	The black dotted line in panel (c) is the linear extrapolation of the frequency dependence of mode $\Phi_1^{(1)}$ in the cycloidal state \cite{Fishman2015}. 
	The gray background denotes the cycloidal state. 
	The yellow background denotes the intermediate spin state between the cycloidal and canted AF states for $\bdc/\parallel \ZZ$ in panel (c). 
	The boundaries are determined from the field dependence of the magnetization $M(B)$ \cite{Tokunaga2014JPSConf}.
	}
\end{figure*} 


\begin{table}
	\caption{Exchange and anisotropy parameters, unit meV,  of \bfo/.
		The spin-wave energies at $\vect{q}=0$ do not depend on the parameters $J_2$ and $D_1$ in the canted AF state.  $K_H$ is hexagonal anisotropy \cite{Fishman2018PhysicaB, Fishman2018} used to model the rhombohedral phase (see Supplemental Material).
		Magnetoelastic anisotropy parameters $K_{E,2}$ and $K_{E,3}$ were determined for the magnetic field directions $\XX$, $\YY$, and $\vect{z}$.
		By symmetry, magnetoelastic anisotropy parameters are zero when $\bdc/\parallel \ZZ$.}
	
	\begin{tabular}{|l|ccc|}
		\hline
		\hline
		& Previous&Method & This work \\
		\hline
		$J_1$  & -5.3&a & -5.3 (fixed)  \\
		$J_2$ & -0.2 &a & - \\
		$D_1$ &0.18 &b& -  \\
		$D_2$ & $6.0\times 10^{-2}$ &c& $(8.32\pm 0.48)\times 10^{-2}$ \\
		$K_Z$ &$4.0\times 10^{-3}$ &d& $(3.76\pm 0.39)\times 10^{-3}$   \\
		$K_H$ & $4 \times 10^{-6}$ &e&  0 \\	
	\hline
		& \multicolumn{3}{c|}{Magnetic field direction (this work)}\\\cline{2-4}
	& \multicolumn{2}{c}{$\XX$} &$\YY, \vect{z}$ \\
		\hline
	$K_{E,2}$ &  \multicolumn{2}{c}{$0$} &$(1.02\pm 0.27)\times 10^{-4}$  \\
	$K_{E,3}$ & \multicolumn{2}{c}{$-(5.90\pm 0.68)\times 10^{-5}$}&  $-(1.14\pm 0.61)\times 10^{-5}$  \\
		\hline
		\hline
	\end{tabular}
	\label{tab:parameters}
	\footnotetext[1]{INS \cite{Jeong2012, Matsuda2012, Xu2012}}
	\footnotetext[2]{Cycloid wavelength \cite{Sosnowska1995} }
	\footnotetext[3]{Cycloid tilt \cite{Tokunaga2010JPSJ, Ramazanoglu2011prl}, INS \cite{Jeong2014}}
	\footnotetext[4]{3rd harmonic generation \cite{Zalesskii2000,Zalesskii2002}, neutron diffraction \cite{Ohoyama2011}, INS \cite{Matsuda2012, Jeong2014}, spectroscopy \cite{Nagel2013, Fishman2013PRB}, tight binding \cite{DeSousa2013}}
	\footnotetext[5]{Cycloid order vector  rotation \cite{Fishman2018PhysicaB, Fishman2018}}
\end{table}

Measured mode frequencies are plotted as a function of magnetic field along $\XX$, $\YY$, $\ZZ$, and cubic axis $\zz$ in Fig.\,\ref{fig:Magnetic_spectra}.  The blue circles and black squares are the spin-wave frequencies.  
The red dashed line gives the linear field dependence of the red circles, which are produced by impurities \cite{SM_BFO}.

The  boundaries between the cycloidal and canted AF state found by THz absorption spectroscopy agree fairly well with the vertical lines in Fig.\,\ref{fig:Magnetic_spectra} obtained from the the maximum of $\mathrm{d}M/\mathrm{d}B$, where $M(B)$ is the magnetization \cite{Tokunaga2014JPSConf,Tokunaga2015a}.  
For field along $\ZZ $, $\mathrm{d}^2M/\mathrm{d} B^2$ vanishes at the upper critical field of the intermediate state, 28\,T.
Scattering of the THz data near 18\,T for field along $\XX$ or $\YY$, Figs.\,\ref{fig:Magnetic_spectra}(a) and (b),
is probably caused by a slight misorientation of the sample relative to $\bdc/$.

At 28\,T, the transition into the canted AF state for $\bdc/\parallel \ZZ$ is clearly marked by the appearance of the $\nu_2$ mode and the disappearance of other modes,  Fig.\,\ref{fig:Magnetic_spectra}(c).
Since strain is absent and there is no in-plane anisotropy for this field direction (we assume $K_H=0$), our model predicts that $\nu_1 = 0$. 

An intermediate spin state appears between the cycloidal and canted AF states when $\bdc/\parallel \ZZ$, Fig.\,\ref{fig:Magnetic_spectra}(c).
In the cycloidal state, the frequency of mode $\Phi_1^{(1)}$ [\onlinecite{Fishman2015}] extrapolates to zero at 24.5\,T, the same field where the cycloidal state transforms to an intermediate state according to magnetization data.
Other modes do not exhibit clear changes when entering this intermediate state.  
Theoretical studies \cite{Gareeva2013} and neutron diffraction spectroscopy \cite{Kawachi2017, Matsuda2020} reveal that the intermediate state in magnetic field $\bdc/\perp \mathbf{Z}$ is a conical spin structure with ordering vector along the magnetic field.  While earlier measurements suggested that it disappears at low $T$ \cite{Kawachi2017},
our data indicate that the intermediate state exists even at low $T$ when $\bdc/\parallel \ZZ$.

Our main theoretical results for the spin-wave frequencies are shown by the solid curves in Fig.\,\ref{fig:Magnetic_spectra}, which were obtained for a magnetoelastically strained crystal using Eq.\,(\ref{eq:ham_fit}).
We introduce ten strain-induced parameters $K_{\Gamma,k}^{(\vect{m})}$:  one set for $\mm=\XX$ and the other set for $\mm=\YY$ or $\zz$.  Recall that strain is absent for $\vect{m}\parallel \vect{Z}$.  Because the 
spins lie in the $XY$ plane, $-K_{E,1}^{(\vect{m})}\sum_i S_{iY}S_{iZ}^3$ does not contribute to the spin dynamics (only two of the three factors of $S_{iZ}$ can be replaced by boson operators in the Holstein-Primakoff expansion \cite{SWBook2018}).
Our fit gave large errors for the parameters $K_{A,1}^{(\vect{m})}$, $K_{A,2}^{(\vect{m})}$, and $K_{E,2}^{(\vect{m}=\vect{X})}$, which were then set to zero.  
Consequently, neither of the $l=2$ monoclinic, single-ion anisotropy terms appear in our Hamiltonian.
Found to be negligible, $K_H$ was also set to zero.
The final fit was then performed with three magnetostriction-enforced parameters together with $D_2$ and $K_Z$:  five parameters in all \footnote{We could have fixed $D_2$
using the measured $M_0$, thereby reducing the number of fitting parameters from five to four.  However, we elected to leave $D_2$ free for two reasons.  First, $M_0$ is not known very accurately from experiment.  
Second, the resulting theoretical value for $M_0$ can be used to test the model.}. 
Table\,\ref{tab:parameters} lists the values for these five parameters.  

Aside from some differences due to the scattering of the experimental points (particularly for $\nu_1$), the agreement between theory and experiment
for the THz frequencies is quite good.
By contrast, the rhombohedral spin model yields a value for $\chi^2$ that is four times larger \cite{SM_BFO}
than our monoclinic model.  

\section{Model parameters}

Table\,\ref{tab:parameters} compares the parameters of the canted AF and cycloidal states. 
While the new estimate for $K_Z$ is close to the previous estimate in the cycloidal state, the new value for $D_2$ is about 39\% larger than the cycloidal estimates.  
 
Our numerical results for the anisotropy parameters agree
with simple estimates based on their order in the spin-orbit
coupling parameter $l|J_1|$ where $l \sim 10^{-1}$.
While the DM interactions are first order in $l$ and the easy-axis anisotropy $K_Z$ is second order in $l$, the magnetoelastic parameters $K_{E,2}$
and $K_{E,3}$ are third order \cite{Fishman2018PhysicaB}.  
Therefore, $S^4 K_{E,n} \sim l S^2 K_Z$ so that 
$K_{E,n} \sim 10^{-2}\, K_Z$, as found in Table\,\ref{tab:parameters}.
Just as inelastic neutron-scattering lacks the energy resolution to determine the  
small DM and anisotropy interactions in \bfo/, it also lacks the energy resolution to determine the even smaller magnetoelastic coupling parameters $K_{E,2}$ and $K_{E,3}$.
Fortunately, the small parameters induced by spin-orbit coupling can be measured using
spectroscopic techniques.

As expected, the canted AF state has a small FM moment in the $XY$ plane induced by the DM interaction $D_2$.
The spin canting and corresponding FM moment $M_0$ can be experimentally estimated by extrapolating the magnetization to zero magnetic field.
While early work \cite{Tokunaga2010} estimated that $M_0 = 0.03$\,$\mu_{\rm B}$ per Fe,  
more recent experiments obtained $M_0=0.048\,\mu_{\rm B}$ \cite{Tokunaga2015} or 0.04\,$\mu_{\rm B}$ \cite{Kawachi2017} per Fe.  

With spins in the $XY$ plane and $K_H=0$, only the $K_{E,3}^{(\vect{ m})}$ term violates rotational invariance.  
The canting angle $\phi_0 \ll 1$ is theoretically given by
\begin{equation}\label{eq:canting_angle}
\phi_0 \approx \frac{1}{2} \frac{D_2}{\vert J_1 \vert + K_{E,3}^{(\vect{m})} S^2/3 }   
\end{equation}
with canted magnetization
\begin{equation}
\label{eq:M0}
M_0 = 2S \mu_{\rm B} \sin \phi_0 \approx \frac{S\mu_{\rm B}D_2}{\vert J_1 \vert + K_{E,3}^{(\vect{m})} S^2/3 }.
\end{equation}
Because $K_{E,3}^{(\vect{m})} S^2/3\vert J_1\vert \sim 10^{-5}$, 
$\phi_0 \approx D_2 /2\vert J_1\vert $
and $M_0 \approx S\mu_{\rm B} D_2 /\vert J_1\vert $ are independent of the direction of the field in the $XY$ plane, in agreement with
the observation that $M_0$ is the same for fields along $\XX $ and $\YY $ \cite{Tokunaga2015}.
The rotational invariance of $M_0$ confirms that magnetostriction affects neither the exchange coupling $J_1$ nor the DM coupling $D_2$:
if $J_1$ or $D_2$ were altered by strain, then $M_0\propto D_2/\vert J_1\vert $ would be different for fields along $\XX$ and $\YY$.
Our result that magnetostriction mostly affects the single-ion anisotropy is consistent with recent {\it ab initio} results that the single-ion anisotropy is highly sensitive to a small misfit of crystal parameters \cite{Chen2018}.

The fitting parameter $D_2 \approx 8.3 \times 10^{-2}$\,meV gives $\phi_0=0.0078 \pm 0.0005$\,rad and $M_0\approx 0.039\pm 0.002$\,$\mu_{\rm B}$ per Fe, which is within range of the two most recent  
experimental estimates \cite{Tokunaga2015, Kawachi2017}. 
By comparison, the value $D_2\approx 6.0 \times 10^{-2}$\,meV obtained from earlier cycloidal state measurements \cite{Tokunaga2010JPSJ, Ramazanoglu2011prl, Jeong2014} and from a rhombohedral fit for the canted AF state \cite{SM_BFO} gives $M_0 = 0.027$\,$\mu_{\rm B}$ per Fe, which is about 33\% smaller than the recent experimental estimate of 0.04\,$\mu_{\rm B}$ \cite{Kawachi2017}. 

\section{Discussion and Conclusion}

Since $K_{E,2}^{(\mm)}$ and $K_{E,3}^{(\mm)}$ depend on the direction $\vect{m}$ of the magnetic field, the strain is different for field along $\XX$ and $\YY$. 
This agrees with the observation that the magnetostriction $\Delta l_X/l_X$ 
at $\vect{B}_{\mathrm{c}}$ \cite{Kawachi2019} is
positive when $ \bdc/ \parallel \XX$ and negative when $ \bdc/ \parallel \YY$.
Moreover, $\Delta l_X/l_X$ is nearly constant as $B$ increases above the critical field. 
Hence, the single-ion contributions of the spin-canting FM component $\vect{S}_1+\vect{S}_2$ 
to the magnetostriction are small compared to the single-ion
contributions of the AF vector $\vect{S}_1-\vect{S}_2$.

Other evidence for magnetoelastic coupling in the canted AF state is provided by the transverse electric polarization $\vect{P}_t\perp \ZZ$, which changes as the magnetic field is rotated in the hexagonal plane \cite{Kawachi2019}.
 If $\vect{m}\parallel \vect{Y}$,  $\vect{P}_t\parallel \vect{Y}$;  if $\vect{m}\parallel \vect{X}$, $\vect{P}_t\parallel -\vect{Y}$.
Both strains $\epsilon_1^{\gamma,1}=[\epsilon_{XX}- \epsilon_{YY} ]/2$ and $\epsilon_1^{\gamma,2}=\epsilon_{YZ}$ preserve the $YZ$ mirror plane and allow  $\vect{P}_t\parallel \vect{Y}$.
Because $\epsilon_{YZ}$ tilts the $Z$ axis, it could produce the in-plane component $\vect{P}_t$ by rotating the FE polarization $\vect{P}$.
A tilting angle of 0.01 to 0.04$^\circ $ is consistent with the magnitude of $P_t$ \cite{Kawachi2017}.

In the cycloidal state,
$\vect{P}_t$ is again modulated by the rotation of an in-plane magnetic field 
with an amplitude roughly half the size of that in the AF state \cite{Kawachi2019}. 
Unlike in the canted AF state, this behavior cannot be explained by the strain $\epsilon_{YZ}$ because a periodic spin structure like the cycloid should not produce homogeneous strain.
Therefore, it is likely \cite{Kawachi2019} that $\vect{P}_t$ is induced by metal-ligand hybridization \cite{Jia2006} and not by the tilting of the $c$ axis in both the cycloidal and AF states.
Additional magnetostriction measurements are needed to determine which strain component,  $\epsilon_1^{\gamma,1}=[\epsilon_{XX}- \epsilon_{YY} ]/2$ or $\epsilon_1^{\gamma,2}=\epsilon_{YZ}$, is dominant in the AF state of \bfo/.

The hysteresis of the magnetostriction \cite{Kawachi2017} and of the cycloidal
wavevector $\vect{Q}$ in a magnetic field \cite{Bordacs2018} also demonstrate that magnetoelastic coupling is important in the cycloidal state.
The rotation of the AF vector $\vect{S}_1-\vect{S}_2$ with the period of the cycloidal wavelength will induce strain at the harmonic wavevectors $2Q$ and $4Q$. 
Consequently, the single-ion anisotropy constants will also be modulated with wavevectors $2Q$ and $4Q$.
However, the spin-wave frequencies of the cycloidal state are (at least so far) well described by the rhombohedral model without additional magnetoelastic couplings.

This work demonstrates that high-resolution THz absorption measurements can be used to determine the magnetoelastic  coupling constants for the AF phase of a material.
The magnetic-field dependence of the $\vq =0 $ spin-wave frequencies in the canted AF state of \bfo/  were fitted using a spin model consistent with the monoclinic distortion of the orthorhombic $R3c$ lattice. 
Whereas epitaxial strain stabilizes the monoclinic phase in thin \bfo/ films \cite{Bea2007}, magnetoelastic coupling stabilizes the monoclinic phase in bulk \bfo/.  
The magnetoelastic coupling is driven by the in-plane spin components parallel to the AF order vector $\vect{S}_1-\vect{S}_2$.  Those spin components couple to the strain through single-ion anisotropy interactions. 
Our new microscopic model for the canted AF state of \bfo/ contains two single-ion terms that appear in monoclinic symmetry and depend on the direction of the magnetic field in the $XY$ plane.  
The dependence of the spin microscopic parameters on the orientation of the magnetic field has clear implications for the technological applications of \bfo/.

Several new questions about bulk \bfo/ are raised by this work.  Density-functional calculations are needed to  understand the  disappearance of the $l=2$ single-ion anisotropy terms.  Magnetostriction measurements are required to  distinguish the strains $\epsilon_1^{\gamma,1}=[\epsilon_{XX}- \epsilon_{YY} ]/2$ and $\epsilon_1^{\gamma,2}=\epsilon_{YZ}$ in the canted AF state.
The appearance of the intermediate conical state for field along $\ZZ$ at low temperatures requires additional study.
New measurements and theory are needed to clarify the role of magnetoelastic coupling in the cycloidal state.
Thus, the proposed model may serve as the foundation for future work on this important multiferroic material, providing 
insight into both the cycloidal and canted AF states.

\begin{acknowledgments}

We thank Bianca Trociewitz  for her contribution to the  design of probes  and  for technical assistance in Tallahassee.
Research  was supported by the European Regional Development Fund project TK134 and by the Estonian Ministry of Education and Research Council Grants IUT23-03 and  PRG736, 
by the bilateral program of the Estonian and Hungarian Academies of Sciences under the Contract No. SNK-64/2013, 
by the Hungarian NKFIH Grants No. K 124176 and ANN 122879, 
by the BME-Nanonotechnology and Materials Science FIKP grant of EMMI (BME FIKP-NAT),
by the FWF Austrian Science Fund I 2816-N27, 
and by the Deutsche Forschungsgemeinschaft (DFG) via the Transregional Research Collaboration TRR 80: From Electronic Correlations to Functionality (Augsburg-Munich-Stuttgart).
T.D. acknowledges funding support from Augusta University and SYSU Grant  OEMT-2017-KF-06 and
R.S.F. by the U.S. Department of Energy, Office of Basic Energy Sciences, Materials Sciences and Engineering Division.
A portion of this work was performed at the National High Magnetic Field Laboratory, which is supported by NSF Cooperative Agreement DMR-1644779 and the State of Florida.
The support of the HFML-RU/FOM, member of the European Magnetic Field Laboratory (EMFL), is acknowledged.

This manuscript has been authored by 
UT-Battelle, LLC under Contract No. DE-AC05-00OR22725 
with the U.S. Department of Energy. 
The United States Government retains and the publisher, by accepting the article for publication, acknowledges that the United States Government retains a non-exclusive, paid-up, irrevocable, world-wide license to publish or reproduce the published form of this manuscript, or allow others to do so, for United States Government purposes. 
The Department of Energy will provide public access to these results of federally sponsored research in accordance with the DOE Public Access Plan \cite{access}.

\end{acknowledgments}
\clearpage

\pagebreak
\widetext
\begin{center}
{\large \textbf{Magnetoelastic Distortion of Multiferroic \bfo/ in the Canted Antiferromagnetic State:  Supplemental Material} \vspace{0.25cm} \\
T. R{\~o}{\~o}m$^{*}$,~J. Viirok,~L. Peedu,~U. Nagel,~D. G. Farkas,~D. Szaller,~V. Kocsis,~S. Bord{\'a}cs,~I. K{\'e}zsm{\'a}rki,~D. L. Kamenskyi,~H. Engelkamp,~M. Ozerov,~D. Smirnov,~J. Krzystek,~K. Thirunavukkuarasu,~Y. Ozaki,~Y. Tomioka,~T. Ito,~T. Datta,~and R. S. Fishman$^{\dagger}$}
\end{center}

\subsection{Theory}
This Supplemental Material develops a new spin model for \bfo/ in the canted AF state.  Magnetoelastic coupling distorts the crystal and breaks rhombohedral symmetry by introducing 
new single-ion anisotropy terms into the spin Hamiltonian. 
The total energy is the sum of elastic, magnetic (spin-only part), and magnetoelastic energies: 
\begin{equation}\label{eq:ham_total}
\mathcal{H} = {\mathcal H}_\mathrm{e}+ {\mathcal H}_\mathrm{m} +  {\mathcal H}_\mathrm{me}.
\end{equation}
To find the equilibrium spin configuration, the energy 
$E = \langle \mathcal{H}\rangle $ is minimized with respect to both the strain and the spin orientations.
The spin-wave spectrum is evaluated using ${\mathcal H}_\mathrm{m}$ plus additional terms arising from ${\mathcal H}_\mathrm{me}$ due to the strain.
Below we write down each part of (\ref{eq:ham_total}), find the strain induced by the zero-order AF spin state, and then construct a new spin Hamiltonian by including additional strain-induced single-ion anisotropy terms.

\subsubsection{Magnetoelastic energy \label{sec:me}}

Our treatment of the magnetoelastic energy follows the microscopic theory of Callen {\it et al.} \cite{Callen1963,Callen1965}. 
The general form for single-ion contributions to the magnetoelastic energy is given by
\begin{equation}\label{eq:H_m-e}
\mathcal{H}_\mathrm{me} = -\sum_{f} \sum_{\Gamma, \Gamma'} \sum_{j,j'}  \tilde{B}_{jj'}^{\Gamma, \Gamma'}(f) \sum_{i} \epsilon_i^{\Gamma,j} \mathcal{S}_i^{\Gamma',j'}\!(f) +\ldots,
\end{equation}
which omits higher-order terms in the strain $\epsilon_i^{\Gamma,j}$.
Since we only consider homogeneous strain, the magnetoelastic energy is the same in each magnetic unit cell, with spins labeled by $f$. 
In \bfo/, the two spins $f=1$ and 2 in the unit cell are equivalent and
the magnetoelastic constants are equal: $\tilde{B}_{jj'}^{\Gamma, \Gamma'}(1)=\tilde{B}_{jj'}^{\Gamma, \Gamma'}(2)$.
Both $\Gamma$ and $\Gamma'$ count the irreducible representations of the spin-site symmetry. 
The final factor in Eq.\,[\ref{eq:H_m-e}], $\mathcal{S}_i^{\Gamma,j}(f)$, is the linear combination of $f$ spin operators that transform like the $i$-th component of the irreducible representation $\Gamma$.
Because the strain is homogeneous, it is sufficient to consider point group symmetry $C_{3v}$ at the spin sites.
If more than one combination of symmetrized strain or spin operators transform like $\Gamma$ ($\Gamma'$), then $j>1$ ($j'>1$).
Index $i$ runs over the components of the basis functions that transform like an $n$-dimensional irreducible representation, $i=1,\ldots, n$. 
The relevant irreducible representations in $C_{3v}$ are $A_1$ and $E$ with $\Gamma=\Gamma'$.

Several criteria constrain the terms in the magnetoelastic Hamiltonian.
Since the Hamiltonian transforms like the fully symmetric representation $A_1$,
spin operators and strain can be combined only when $\Gamma \otimes \Gamma'$ contains $A_1$.
The strain is time-inversion invariant but the spin is not. 
Because the Hamiltonian must be invariant with respect to time inversion, only even powers of spin operators are allowed in $\mathcal{S}_i^{\Gamma,j}$.

Our symmetry analysis is simplified by expanding
the strain and spin tensors in spherical harmonics  $\SpherHar{l}{m}$ and $\SpherOp{l'}{m'}$, respectively.
The strain tensor only has $l=0$ and $l=2$ components.
Even powers of spin operators require even values of $l'$ 
and we limit $l'$ to values of 2 and 4.

Following the symmetry analysis in Section \ref{sec:sym_me_C3v}, Table\,\ref{tab:l2_l4} presents the $A_1$-symmetric magnetoelastic terms of Eq.\,\ref{eq:H_m-e} that are linear in strain and second or fourth order in the spin.
The magnetoelastic constants $\tilde{B}_{jj'}^{\alpha}$ and $\tilde{B}_{jj'}^{\gamma}$ are explained in Table\,\ref{tab:Spher_Cart_C3v}, which provides the mapping between tensors in spherical $\{\theta,\phi\}$ and Cartesian  coordinates $\{ x,y,z\}$.
Since the spin is treated classically, this mapping also applies to the spin 
(for quantum-mechanical spins, the mapping of tensor operators from spherical to Cartesian coordinates 
was given by Ref.\,[\onlinecite{Callen1963},\onlinecite{Zare1988}]).
Labels $\alpha$ and $\gamma $ denote tensors of $A_1$ and $E$ symmetry, respectively.
In $C_{3v}$, $\Gamma \otimes \Gamma'$ contains $A_1$ only if $\Gamma = \Gamma'$.

Altogether, four spherical tensor functions transform like the irreducible representation $A_1$ and five transform like the two-dimensional irreducible representation $E$ for $l=0, 2$ and 4.
For example, $\tilde{B}_{23}^{\gamma}$ is the coupling between strain represented by the spherical tensor function $(\gamma, 2)$ and the spin function represented by the spherical tensor function $(\gamma, 3)$, Table\,\ref{tab:Spher_Cart_C3v}.
Because \SpherOp{0}{0} term produces a constant offset of the energy ($S_x^2 + S_y^2 + S_z^2 = S^2$), it is not included in $\mathcal{H}_\mathrm{me}$. 

\begin{table}
	\caption{	\label{tab:l2_l4} $A_1$-symmetric magnetoelastic coupling terms between the spherical strain tensor $\SpherHar{l}{m}$ ($l=0,2$) and 
	the spherical  spin  tensor $\SpherOp{l}{m}$ ($l=2,4$) in  point group $C_{3v}$. 
	$\ket{\SpherHar{l}{m}}_+ = \SpherHar{l}{m}  + \SpherHar{l}{-m} $ and 	$\ket{\SpherHar{l}{m}}_- = \SpherHar{l}{m}  - \SpherHar{l}{-m}$ where $0\leq m\leq l$.
	$\SpherHar{l}{m}(\theta,\phi)$ are the basis functions of a full rotation group $O(3)$ as defined in Ref.\,[\onlinecite{Altmann2011}].}
	\begin{tabular}{c|l}
		\hline	\hline
		Coupling constant & Spherical harmonics\\
		\hline
		$\tilde{B}_{12}^{\alpha}$	 & $\SpherHar{0}{0} \SpherOp{2}{0}$ \\
		$\tilde{B}_{22}^{\alpha}$	 & $\SpherHar{2}{0} \SpherOp{2}{0}$ \\
		$\tilde{B}_{23}^{\alpha}$	 & $\SpherHar{2}{0} \SpherOp{4}{0}$ \\
		$\tilde{B}_{14}^{\alpha}$	& $\SpherHar{0}{0} \ket{\SpherOp{4}{3}}_+$ \\
		$\tilde{B}_{24}^{\alpha}$	& $\SpherHar{2}{0} \ket{\SpherOp{4}{3}}_+$ \\
		$\tilde{B}_{11}^{\gamma}$	&$\ket{\SpherHar{2}{2}}_+ \ket{\SpherOp{2}{2}}_+ - \ket{\SpherHar{2}{2}}_- \ket{\SpherOp{2}{2}}_-$\\
		$\tilde{B}_{12}^{\gamma}$	&$\ket{\SpherHar{2}{2}}_+ \ket{\SpherOp{2}{1}}_+ + \ket{\SpherHar{2}{2}}_- \ket{\SpherOp{2}{1}}_-$\\
		$\tilde{B}_{13}^{\gamma}$	&$\ket{\SpherHar{2}{2}}_+ \ket{\SpherOp{4}{1}}_+ + \ket{\SpherHar{2}{2}}_- \ket{\SpherOp{4}{1}}_-$\\
		$\tilde{B}_{14}^{\gamma}$	&$\ket{\SpherHar{2}{2}}_+ \ket{\SpherOp{4}{2}}_+ - \ket{\SpherHar{2}{2}}_- \ket{\SpherOp{4}{2}}_-$\\
		$\tilde{B}_{15}^{\gamma}$	&$\ket{\SpherHar{2}{2}}_+ \ket{\SpherOp{4}{4}}_+ + \ket{\SpherHar{2}{2}}_- \ket{\SpherOp{4}{4}}_-$\\
		$\tilde{B}_{21}^{\gamma}$	&$\ket{\SpherHar{2}{1}}_+ \ket{\SpherOp{2}{2}}_+ + \ket{\SpherHar{2}{1}}_- \ket{\SpherOp{2}{2}}_-$\\
		$\tilde{B}_{22}^{\gamma}$	&$\ket{\SpherHar{2}{1}}_+ \ket{\SpherOp{2}{1}}_+ - \ket{\SpherHar{2}{1}}_- \ket{\SpherOp{2}{1}}_-$\\
		$\tilde{B}_{23}^{\gamma}$	&$\ket{\SpherHar{2}{1}}_+ \ket{\SpherOp{4}{1}}_+ - \ket{\SpherHar{2}{1}}_- \ket{\SpherOp{4}{1}}_-$\\
		$\tilde{B}_{24}^{\gamma}$	&$\ket{\SpherHar{2}{1}}_+ \ket{\SpherOp{4}{2}}_+ + \ket{\SpherHar{2}{1}}_- \ket{\SpherOp{4}{2}}_-$\\
		$\tilde{B}_{25}^{\gamma}$	&$\ket{\SpherHar{2}{1}}_+ \ket{\SpherOp{4}{4}}_+ - \ket{\SpherHar{2}{1}}_- \ket{\SpherOp{4}{4}}_-$\\
		\hline
		\hline
	\end{tabular}
\end{table}

\begin{table}
	\caption{Mapping between  tensors  in spherical $\{\theta,\phi\}$ and Cartesian $\{ x,y,z\}$ coordinates for the irreducible representations  $A_1$ and $E$ in $C_{3v}$ for $l=0, 2, 4$.
	$\ket{\SpherHar{l}{m}}_+ = \SpherHar{l}{m}  + \SpherHar{l}{-m} $ and 	$\ket{\SpherHar{l}{m}}_- = \SpherHar{l}{m}  - \SpherHar{l}{-m}$ where $0\leq m\leq l$.
	$\SpherHar{l}{m}(\theta,\phi)$ are the basis functions of the full rotation group $O(3)$ as defined in Ref.\,[\onlinecite{Altmann2011}].
	The same mapping applies between the spherical spin tensor components  and the Cartesian  spin tensor components for a classical spin.
	}
	\label{tab:Spher_Cart_C3v}
	\begin{tabular}{cc|c|c}\hline\hline
		Irrep&  & Spherical & Cartesian  \\ \hline
		$A_1$ & $\alpha,1$ & $\SpherHar{0}{0}$ &1\\
		$A_1$ &$\alpha,2$& $\SpherHar{2}{0}$ & $z^2$\\
		$A_1$ &$\alpha,3$& $\SpherHar{4}{0}$ & $z^4$\\
		$A_1$ &$\alpha,4$& $i\ket{\SpherHar{4}{3}}_+$ & ($3x^2 - y^2) y z$\\
		$E$ &$\gamma,1$ & $\{\ket{\SpherHar{2}{2}}_+, -i\ket{\SpherHar{2}{2}}_- \}$ & $\{\frac{1}{2}(x^2-y^2), xy \}$ \\
		$E$ & $\gamma,2$ & $\{i\ket{\SpherHar{2}{1}}_+, -\ket{\SpherHar{2}{1}}_- \}$ & $\{y z, x z \}$  \\
		$E$ &$\gamma,3$ & $\{i\ket{\SpherHar{4}{1}}_+, -\ket{\SpherHar{4}{1}}_- \}$ & $\{y z^3, x z^3\}$\\
		$E$ &$\gamma,4$ & $\{\ket{\SpherHar{4}{2}}_+, -i\ket{\SpherHar{4}{2}}_- \}$ & $\{\frac{1}{2}(x^2z^2-y^2z^2), xyz^2 \}$\\
		$E$ &$\gamma,5$ & $\{\ket{\SpherHar{4}{4}}_+,  -i\ket{\SpherHar{4}{4}}_- \}$ & $\{\frac{1}{4}(x^4 + y^4 - 6 x^2y^2), x^3y- xy^3 \}$\\
		\hline\hline
	\end{tabular}
\end{table}

\subsubsection{Elastic energy \label{sec:elastic_energy}}

The general form for the elastic energy is given by \cite{Callen1965}
\begin{equation}\label{eq:H_elastic}
\mathcal{H}_e = \sum_{\Gamma, \Gamma'} \sum_{j,j'} \frac{1}{2} c_{jj'}^{\Gamma,\Gamma'} \sum_{i} \epsilon_i^{\Gamma,j} \epsilon_i^{\Gamma',j'},
\end{equation}
where $\epsilon_i^{\Gamma,j}$ is an $n$-dimensional $(i=1,\ldots, n$) strain function corresponding to representation $\Gamma$,
$j$ counts the strain functions if there are more than one for a given representation $\Gamma$, and
$c_{jj'}^\Gamma $ are the elastic constants.

Symmetrized strain tensors for the point group $C_{3v}$ are \cite{Callen1965}
\begin{eqnarray}
\epsilon^{\alpha,1} &= & \epsilon_ {XX} + \epsilon_ {YY} + \epsilon_ {ZZ}, \label{eq:strain_A11} \\
\epsilon^{\alpha,2} &= & \frac{1}{2}\Bigl(\sqrt{3} \epsilon_{ZZ} - \frac{1}{\sqrt{3}}\epsilon^{\alpha,1}\Bigr), \label{eq:strain_A12}\\
\{\epsilon_1^{\gamma, 1}, \epsilon_2^{\gamma, 1}\} & = & \Bigl\{ \frac{1}{2}(\epsilon_{XX}- \epsilon_{YY} ), \epsilon_{XY} \Bigr\}, \label{eq:strain_Egamma}\\
\{\epsilon_1^{\gamma, 2 }, \epsilon_2^{\gamma, 2}\} & = & \{\epsilon_{YZ},  \epsilon_{XZ}\}. \label{eq:strain_Eepsilon}
\end{eqnarray}
Using Table\,\ref{tab:Spher_Cart_C3v} and the requirement that $\Gamma \otimes \Gamma'$ must contain $A_1$, the $C_{3v}$-symmetric elastic energy is 
\begin{eqnarray} 
&&\mathcal{H}_{\mathrm{e}}= \frac{1}{2}c_{11}^\alpha (\epsilon^{\alpha,1})^2+ 
c_{12}^\alpha \epsilon^{\alpha,1}\epsilon^{\alpha,2}+
\frac{1}{2}c_{22}^\alpha (\epsilon^{\alpha,2})^2  \label{eq:e}\\
&&+  \frac{1}{2} c_{11}^{\gamma} [(\epsilon_1^{\gamma,1})^2  +(\epsilon_2^{\gamma,1})^2 ]\label{eq:e_gamma11} +   c_{12}^{\gamma} [\epsilon_1^{\gamma,1} \epsilon_1^{\gamma,2} + \epsilon_2^{\gamma,1} \epsilon_2^{\gamma,2}] \label{eq:e_gamma12} \nonumber \\
&&+  \frac{1}{2} c_{22}^{\gamma} [(\epsilon_1^{\gamma,2})^2  +(\epsilon_2^{\gamma,2})^2 ]\label{eq:e_gamma22}. \nonumber
\end{eqnarray}

\subsubsection{Magnetic energy \label{sec:magn_energy}}

Within rhombohedral symmetry, the magnetic Hamiltonian of the AF state (see main text) is  
\begin{eqnarray}
\label{Ham_magn_AF}
&&{\mathcal H}_\mathrm{m}^{\mathrm{AF}} = -J_1\sum_{\langle i,j\rangle }\vS_i\cdot \vS_j \\
&& + D_2 \, \sum_{\langle i,j\rangle } \, (-1)^\hi \,\zp \cdot  (\vS_i\times\vS_j) -K_Z \sum_i  S_{iZ}^2 \nonumber \\ 
&& -\frac{1}{2}K_H \sum_i \left[(S_{iX}+i S_{iY})^6 + (S_{iX} - i S_{iY})^6\right] \nonumber \\
&&- g\mb B \sum_i \vm \cdot \vS_i.\nonumber 
\end{eqnarray}
Because the hexagonal anisotropy is very weak ($K_H S^6 \ll g\mu_{\rm B} BS$), the magnetic field determines the zero-order spin state discussed next.

\subsubsection{Strain induced by zero-order AF spin  state  \label{sec:strain}}

Based on Eq.\,(\ref{Ham_magn_AF}), we solve for the strain using an approximate ``zero-order'' spin state.
Neglecting the small canting induced by the \DM/ interaction and magnetic field, $\vect{S}_1\approx -\vect{S}_2 $ with both spins perpendicular to the applied magnetic field.
Since $\vert D_2 \vert \gg K_Z$, the spins are forced into the hexagonal plane with $S_{iZ}\approx 0$.
Hence, the lowest-order magnetoelastic energy does not contain $S_Z$ terms for fields along $\XX $ or $\YY $.

Using symmetry-allowed terms from Table\,\ref{tab:l2_l4} and the mapping between tensors in spherical and Cartesian coordinates from Table\,\ref{tab:Spher_Cart_C3v}, the magnetoelastic coupling of Eq.\,(\ref{eq:H_m-e}) for one spin is 
\begin{eqnarray}\label{eq:ham_me_Sz0}
&&\mathcal{H}_\mathrm{me}^i= 
-\frac{1}{2}[ \tilde{B}_{11}^{\gamma} \epsilon_1^{\gamma, 1} +\tilde{B}_{21}^{\gamma}\epsilon_1^{\gamma, 2} ] (S_{iX}^2-S_{iY}^2)\\
&&- \frac{1}{4}[\tilde{B}_{15}^{\gamma} \epsilon_1^{\gamma, 1} +\tilde{B}_{25}^{\gamma}\epsilon_1^{\gamma, 2} ] (S_{iX}^4 + S_{iY}^4 - 6 S_{iX}^2 S_{iY}^2).\nonumber
\end{eqnarray}
Thus, the number of strain tensor components is reduced from six to two with $\epsilon_1^{\gamma,1}=\frac{1}{2}[\epsilon_{XX}- \epsilon_{YY} ]$  and $\epsilon_1^{\gamma,2}=\epsilon_{YZ}$, Table\,\ref{tab:Spher_Cart_C3v}.

Assuming the strain is weak enough that $S_Z\approx 0$ and $\vect{S}_1\approx -\vect{S}_2 $, the equilibrium strain is obtained from the minimization conditions
\begin{eqnarray}\label{eq:H_e_me_minimizeOne}
\frac{\partial}{\partial \epsilon_1^{\gamma,1}}[  \mathcal{H}_{\mathrm{e}}+\mathcal{H}_{\mathrm{me}}^1+\mathcal{H}_{\mathrm{me}}^2] = 0,\\
\frac{\partial}{\partial \epsilon_1^{\gamma,2}} [  \mathcal{H}_{\mathrm{e}}+\mathcal{H}_{\mathrm{me}}^1+\mathcal{H}_{\mathrm{me}}^2] = 0,\label{eq:H_e_me_minimizeTwo}
\end{eqnarray}
where the magnetoelastic energy for spins $S_1$ and $S_2$ is given by (\ref{eq:ham_me_Sz0}).
Only even powers of the spin contribute to (\ref{eq:ham_me_Sz0}) and the magnetoelastic couplings are equal for the two spins.

Spin-induced strain $\epsilon_i^{\Gamma,j}(\vect{m})$ depends on the direction of the magnetic field $\vect{m}$.  We consider four field directions, $\XX$,  $\YY$, $\ZZ$, and $\vect{z}$, Fig.\,1 in paper.
If $\vect{m}= \ZZ$ there is no preferred spin orientation in the hexagonal $XY$ plane so that $\epsilon_1^{\gamma,1}(Z)=\epsilon_1^{\gamma,2}(Z)=0$.
If $\vect{m}= \XX$, then $\vect{S}_1=-\vect{S}_2 =S \YY$ and if  $\vect{m}= \YY$, then $\vect{S}_1=-\vect{S}_2 =S \XX$.
If $\vect{m}=\vect{z}$, the spins are along the $X$ direction with the same spin state as for $\vect{m} =\YY $.
Solving Eqs.\,(\ref{eq:H_e_me_minimizeOne}) and (\ref{eq:H_e_me_minimizeTwo}), we find:
\begin{eqnarray}
{\epsilon}_1^{\gamma,1}(X) &=& c^{-1}(2b_1 S^2+b_2 S^4), \\ 
{\epsilon}_1^{\gamma,2}(X) &=& c^{-1}(2b_3S^2+b_4 S^4), \\
{\epsilon}_1^{\gamma,1}(Y) &=& c^{-1}(-2b_1S^2+b_2 S^4), \\ 
{\epsilon}_1^{\gamma,2}(Y) &=& c^{-1}(-2b_3S^2+b_4 S^4),
\end{eqnarray}
where 
\begin{eqnarray}
b_1 &=& \tilde{B}_{11}^{\gamma} c_{22}^\gamma-\tilde{B}_{21}^{\gamma} c_{12}^\gamma, \\ 
b_2 &=& \tilde{B}_{25}^{\gamma} c_{12}^\gamma - \tilde{B}_{15}^{\gamma} c_{22}^\gamma, \\
b_3 &=& \tilde{B}_{21}^{\gamma} c_{11}^\gamma-\tilde{B}_{11}^{\gamma} c_{12}^\gamma, \\ 
b_4 &=& \tilde{B}_{15}^{\gamma} c_{12}^\gamma - \tilde{B}_{25}^{\gamma} c_{11}^\gamma, \\
c &=& 2[ (c_{12}^\gamma)^2-c_{11}^\gamma c_{22}^\gamma].
\end{eqnarray}

\subsubsection{Hamiltonian with ``frozen'' strain}

The Hamiltonian is the sum of ${\mathcal H}_\mathrm{m}^{\mathrm{AF}}$ from Eq.\,(\ref{Ham_magn_AF}) and the single-ion magnetoelastic interactions: 
\begin{equation}
{\mathcal H}_\mathrm{m}^{\mathrm{SW}}={\mathcal H}_\mathrm{m}^{\mathrm{AF}} + \sum_{i} \mathcal{H}_\mathrm{me}^i(\vect{m}).
\end{equation}
The magnetoelastic Hamiltonian for the $i$-th spin is
\begin{eqnarray}\label{eq:me_short}
	&&\mathcal{H}_\mathrm{me}^i(\vect{m}) = -K_{A,1}^{(\vect{m})} S_{iY} S_{iZ}
	-K_{A,2}^{(\vect{m})} (S_{iX}^2-S_{iY}^2)\\
	&&-K_{E,1}^{(\vect{m})} S_{iY} S_{iZ}^3
	-K_{E,2}^{(\vect{m})} (S_{iX}^2-S_{iY}^2)S_{iZ}^2\nonumber\\
	&&-K_{E,3}^{(\vect{m})} (S_{iX}^4 + S_{iY}^4 - 6 S_{iX}^2 S_{iY}^2).\nonumber
\end{eqnarray}
The single-ion anisotropy constants depend on the {\em direction}-dependent strain found in Section\,\ref{sec:strain}:
\begin{eqnarray}
	K_{A,1}^{(\vect{m})} &=& [\tilde{B}_{12}^{\gamma}\, \epsilon_1^{\gamma, 1}(\vect{m}) +\tilde{B}_{22}^{\gamma}\,\epsilon_1^{\gamma, 2}(\vect{m}) ],\label{eq:SIA_me} \\
	K_{A,2}^{(\vect{m})}&=& \frac{1}{2} [ \tilde{B}_{11}^{\gamma} \,\epsilon_1^{\gamma, 1}(\vect{m}) +\tilde{B}_{21}^{\gamma}\,\epsilon_1^{\gamma, 2}(\vect{m}) ], \\
	K_{E,1}^{(\vect{m})} &=&[ \tilde{B}_{13}^{\gamma} \,\epsilon_1^{\gamma, 1}(\vect{m}) +\tilde{B}_{23}^{\gamma}\,\epsilon_1^{\gamma, 2}(\vect{m}) ],\\
	K_{E,2}^{(\vect{m})}&=&  \frac{1}{2}[\tilde{B}_{14}^{\gamma} \,\epsilon_1^{\gamma, 1}(\vect{m}) +\tilde{B}_{24}^{\gamma}\,\epsilon_1^{\gamma, 2}(\vect{m}) ], \\
	K_{E,3}^{(\vect{m})}&=& \frac{1}{4}[\tilde{B}_{15}^{\gamma} \,\epsilon_1^{\gamma, 1}(\vect{m}) +\tilde{B}_{25}^{\gamma}\, \epsilon_1^{\gamma, 2}(\vect{m}) ].
\end{eqnarray}
As shown by Eq,\,(\ref{eq:ham_me_Sz0}), only strains $\epsilon_1^{\gamma,1}=[\epsilon_{XX}- \epsilon_{YY} ]/2$  and $\epsilon_1^{\gamma,2}=\epsilon_{YZ}$ couple to the zero-order spin state.
Hence, only five out of the fifteen magnetoelastic couplings listed in Table\,\ref{tab:l2_l4} survive.


\subsubsection{Symmetry considerations  of magnetoelastic   terms in $C_{3v}$\label{sec:sym_me_C3v}}

Point group $C_{3v}$ has three classes of symmetry elements, $\{E, 2 C_3, 3\sigma_v\}$, and three irreducible representations, $\{A_1, A_2, E\}$.
The $C_3$ elements are $\pm 2\pi/3$ rotations about the $Z$ axis,  Fig.\,1 in paper.
The vertical reflection plane $\sigma_v^X$ is the $YZ$ plane normal to $\XX$.  Two mirror planes are generated by $\pm 2\pi/3$ rotations of $\sigma_v^X$.
It is once again helpful to expand the strain and spin tensors in spherical harmonics \cite{Callen1965}. 
The subduction from the full rotation group $O(3)$ to the point group $C_{3v}$ is given in Table\,\ref{tab:spher_rep_C3v} up to $l=4$.

\begin{table}
	\caption{Irreducible representations of spherical harmonics \SpherHar{l}{m} in the point group $C_{3v}$ [\onlinecite{Altmann2011}]. }
	\label{tab:spher_rep_C3v}
	\begin{tabular}{c|ccc}\hline\hline
		$l$ & $A_1$ & $A_2$ & $E$ \\ \hline
		0&	1 & 0 & 0 \\
		1&	1 & 0 & 1 \\
		2&	1 & 0 & 2 \\
		3&	2 & 1 & 2 \\
		4&	2 & 1 & 3 \\
		\hline\hline
	\end{tabular}
\end{table}

\begin{table}
	\caption{Product of irreducible representations of  $C_{3v}$. }
	\label{tab:product_irrep_C3v}
	\begin{tabular}{c|ccc}\hline\hline
		& $A_1$ & $A_2$ & $E$ \\ \hline
		$A_1$&	$A_1$ & $A_2$ & $E$ \\
		$A_2$&	$A_2$ & $A_1$ & $E$ \\
		$E$&	$E$ & $E$ & $A_1 \oplus A_2 \oplus E$ \\
		\hline\hline
	\end{tabular}
\end{table}

The strain and spin parts of the magnetoelastic Hamiltonian are represented by spherical harmonics of even $l$.
Table \ref{tab:spher_rep_C3v} shows that $l=0$ transforms like $\Gamma^{(0)}  =  A_1 $, $l=2$ like $\Gamma^{(2)}  =  A_1  \oplus 2 E$, and $l=4$ like   $\Gamma^{(4)}  = 2 A_1 \oplus A_2 \oplus 3 E$.
The magnetoelastic Hamiltonian must be fully $A_1$ symmetric.
Table \ref{tab:product_irrep_C3v} implies that the fully symmetric $A_1$ is present in $A_1 \otimes A_1$, $A_2 \otimes A_2$, and $E \otimes E$.
Since the strain is even in $l$, $A_2$ terms are excluded.
Therefore, we are left with combinations of strain and spin parts that are either $A_1$ or $E$.

Table\,\ref{tab:Spher_Cart_C3v} lists $A_1-$ and $E$-symmetric spherical tensors and their mapping to Cartesian tensors.
The $A_1$-symmetric terms in the magnetoelastic coupling (\ref{eq:H_m-e}) are listed in Table\,\ref{tab:l2_l4}.
The $A_1$ symmetry of $E\otimes E$ terms was checked by applying $C_3$ and $\sigma_v^X$ symmetry operations.
Under rotation, the transformation of the strain tensor components $\SpherHar{l}{m}$ and spin tensor components $\SpherOp{l}{m}$ is identical: $C_3 \SpherHar{l}{m} =e^{i m 2\pi/3}\SpherHar{l}{m}$ and  $C_3\SpherOp{l}{m} =e^{i m 2\pi/3}\SpherOp{l}{m}$.
Their transformation is different under reflection: $\sigma_v^X \SpherHar{l}{m}= \SpherHar{l}{-m}$ and $\sigma_v^X \SpherOp{l}{m}= (-1)^l\SpherOp{l}{-m}$.

\begin{table}
	\caption{\label{tab:strained_me2} Spin terms that couple to strains 
	$\ket{\SpherHar{2}{2}}_+=\epsilon_1^{\gamma,1}$  and $\ket{i\SpherHar{2}{1}}_+ = \epsilon_1^{\gamma,2}$. 
	Compared to Table\,\ref{tab:l2_l4}, terms not containing $\ket{\SpherHar{2}{2}}_+$ or $\ket{\SpherHar{2}{1}}_+$ are absent.
	}
	\begin{tabular}{c|r|r}
		\hline	\hline
		$\tilde{B}_{12}^{\gamma}$	&$\ket{\SpherHar{2}{2}}_+ \ket{\SpherOp{2}{1}}_+ $&$\epsilon_1^{\gamma,1}  S_Y S_Z $\\
		$\tilde{B}_{22}^{\gamma}$	&$  \ket{\SpherHar{2}{1}}_+ \ket{\SpherOp{2}{1}}_+$&$ -\epsilon_1^{\gamma,2} S_YS_Z$\\
		$\tilde{B}_{11}^{\gamma}$	&$\ket{\SpherHar{2}{2}}_+ \ket{\SpherOp{2}{2}}_+ $&$\frac{1}{2}\epsilon_1^{\gamma,1}  (S_X^2-S_Y^2) $\\
		$\tilde{B}_{21}^{\gamma}$	&$  \ket{\SpherHar{2}{1}}_+ \ket{\SpherOp{2}{2}}_+$&$ \frac{1}{2} \epsilon_1^{\gamma,2}(S_X^2-S_Y^2)$\\
		$\tilde{B}_{13}^{\gamma}$	&$\ket{\SpherHar{2}{2}}_+ \ket{\SpherOp{4}{1}}_+$&$\epsilon_1^{\gamma,1} S_Y S_Z^3$\\
		$\tilde{B}_{23}^{\gamma}$	&$  \ket{\SpherHar{2}{1}}_+ \ket{\SpherOp{4}{1}}_+$&$ -\epsilon_1^{\gamma,2} S_Y S_Z^3$\\
		$\tilde{B}_{14}^{\gamma}$	&$\ket{\SpherHar{2}{2}}_+ \ket{\SpherOp{4}{2}}_+ $&$\frac{1}{2}\epsilon_1^{\gamma,1} (S_X^2 -S_Y^2 )S_Z^2 $\\
		$\tilde{B}_{24}^{\gamma}$	&$ \ket{\SpherHar{2}{1}}_+ \ket{\SpherOp{4}{2}}_+$&$ \frac{1}{2}\epsilon_1^{\gamma,2} (S_X^2 -S_Y^2 )S_Z^2$\\
		$\tilde{B}_{15}^{\gamma}$	&$\ket{\SpherHar{2}{2}}_+ \ket{\SpherOp{4}{4}}_+ $&$\frac{1}{4}\epsilon_1^{\gamma,1} (S_X^4 + S_Y^4 - 6 S_X^2 S_Y^2) $\\
		$\tilde{B}_{25}^{\gamma}$	&$  \ket{\SpherHar{2}{1}}_+ \ket{\SpherOp{4}{4}}_+$&$ \frac{1}{4}\epsilon_1^{\gamma,2}(S_X^4 + S_Y^4 - 6 S_X^2 S_Y^2)$\\
		\hline
		\hline
	\end{tabular}
\end{table}

\subsection{The Spin Model of the Canted Antiferromagnetic State in Rhombohedral Symmetry}

\begin{figure*}[!thbp]
	\includegraphics[width=1\textwidth]{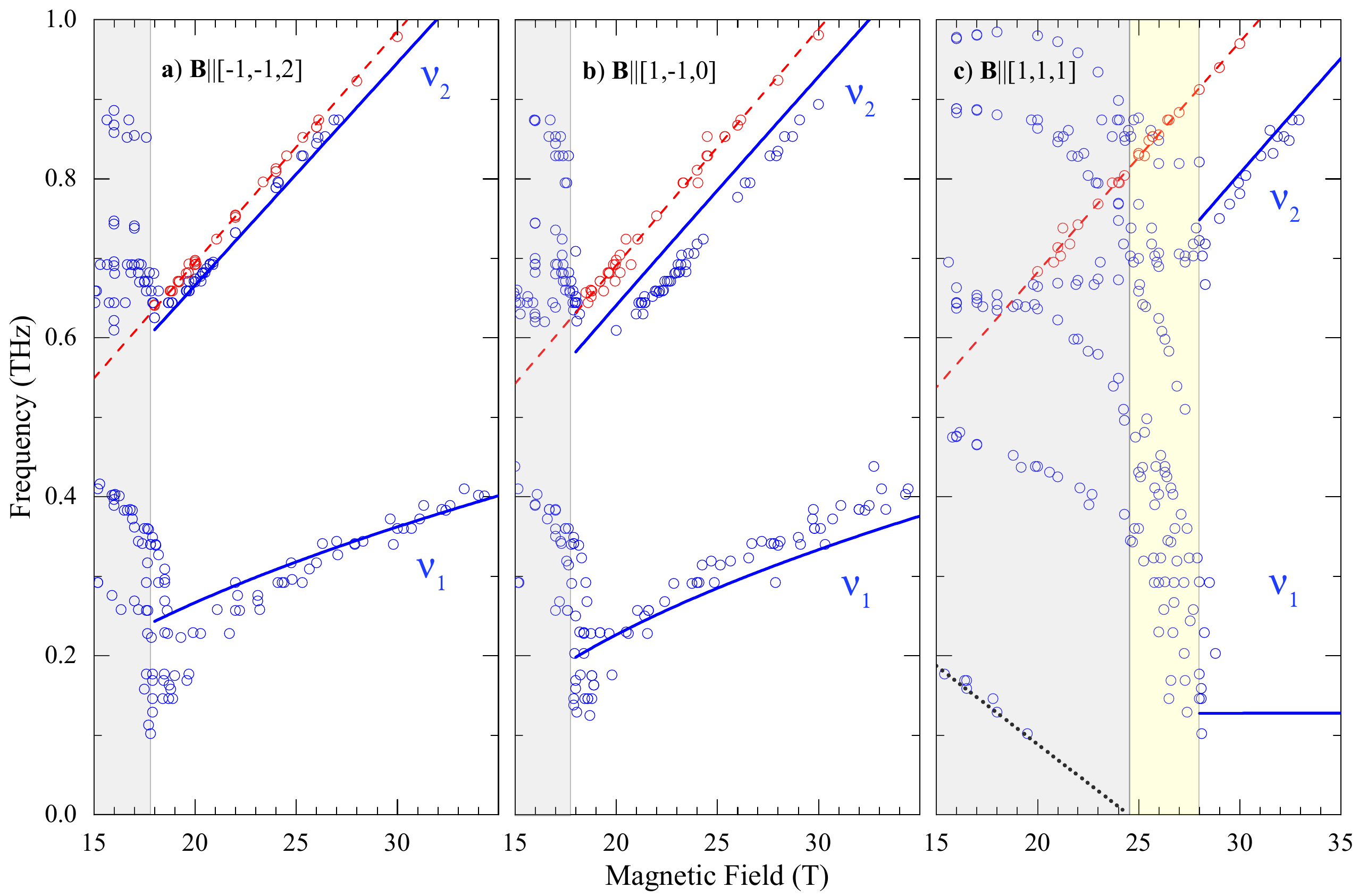}  
	\caption{\label{fig:fit_R3c} The magnetic field dependence of the spin-wave frequencies for field directions (a) $\vect{Y}=[-1,-1,2]$,  (b) $\vect{X}=[1,-1,0]$, and  (c) $\vect{Z}=[1,1,1]$ at liquid He temperature.
		Experimental points are marked by blue circles. 
		Solid blue lines are  the best fit results for spin-wave modes $\nu_1$ and $\nu_2$ using the model spin Hamiltonian (\ref{Ham_magn_AF}) consistent with rhombohedral $R3c$ symmetry.}
\end{figure*} 

The spin model for \bfo/ with rhombohedral symmetry in the canted AF state is given by Eq.\,[\ref{Ham_magn_AF}] with no additional magnetoelastic terms.

Figure \ref{fig:fit_R3c} shows our fit to this rhombohedral model. 
The predicted mode frequencies do not follow the experimental data for field along $\vect{X}$ or  $\vect{Y}$, panels (a) and (b).  Mode $\nu_2$ frequencies are closer to the experimental values for field along $\vect{Z}$, panel (c), but there are still deviations just above $B_c$.
 The frequency of mode $\nu_1$ is nonzero when $\bdc/\parallel \ZZ$ because the hexagonal anisotropy $K_H$ breaks the rotational invariance in the $XY$ plane.

For this rhombohedral fit, $\chi^2 $ is about four times larger than for the monoclinic fit discussed in the main paper.  As described in the main paper,
 $J_1=-5.3$\,meV was taken from inelastic-neutron scattering measurements \cite{Jeong2012, Matsuda2012, Xu2012}.
The rhombohedral parameters are then $D_2=(5.7 \pm 0.6)\e{-2}$\,meV,  $K_Z=(2.1 \pm 0.5)\e{-3}$\,meV, and $K_H=(1.3 \pm 0.4)\e{-6}$\,meV.  
 While the rhombohedral value for $D_2$ is consistent with the cycloidal value (see main paper), $K_Z$ is about half as large.
 The magnetic moment extrapolated to zero field is almost the same for field along $\XX$ and $\YY$: $M_0=S\mu_{\rm B}D_2/\vert J_1\vert \approx 0.027\pm 0.003\,\mu_{\rm B}$ per Fe. 
 This is 33\% less than the lower experimental estimate \cite{Kawachi2017} of 0.04\,$\mu_{\rm B}$ per Fe.
In the absence of the hexagonal anisotropy $K_H$, $\chi^2$ for the rhombohedral model would increase by another factor of two.

\subsection{Impurity mode}
The mode plotted by the red circles and fitted with a linear field dependence (the red dashed line) in Fig.\,\ref{fig:fit_R3c} is assigned to impurities for three reasons. 
First, this mode was absent in flux-grown crystals \cite{Nagel2013}.
Second, the frequencies of this mode do not depend on field orientation.  
Finally, the spin-wave model with two spins in the unit cell permits only two modes.
One candidate for the impurity is Fe in the low spin $S=1/2$ state \cite{PradoGonjal2011}, which is 
insensitive to single-ion anisotropies.  If the orbital moment of the impurity is quenched ($L=0$), then the impurity mode would depend isotropically on the magnetic field, as observed.
The average of the impurity spin parameters in three magnetic field directions gives the $g$ factor $g=2.11 \pm 0.04$ and a zero field intercept of $90 \pm 2$\,GHz.

Due to the distribution of local fields produced by the iron spins,
the impurity signal does not appear in the cycloidal state below 18\,T. 
Therefore, the impurities do not constitute a separate phase with a different structure or chemical composition.  Rather, they are randomly distributed within \bfo/.

\subsection{Experimental Methods}
\bfo/ crystals were grown by the floating zone method using laser diodes as the heat source \cite{Ito2011}. 
Samples in three hexagonal orientations with large faces normal to \mbox{[1,-1,0]}, \mbox{[-1,-1,2]} and \mbox{[1,1,1]} were cut to a thickness of about 0.5\,mm.

THz absorption measurements used either Fourier transform far-infrared (FIR) or continuous wave (CW) spectroscopy.
FIR measurements were performed above 0.55\,THz with a Genzel-type interferometer (Bruker 113v) and a 1.6\,K composite Si bolometer (Infrared Laboratories) as a detector. The radiation source was a mercury arc lamp.
The spectra were collected in a fixed magnetic field.
CW measurements were performed at a fixed frequency by sweeping the magnetic field, a method also called sub-millimeter wave ESR. 
Monochromatic radiation was provided by few frequency-tunable backward wave oscillators and frequency multipliers covering 0.1 to 0.9 THz. 
The radiation intensity was measured with a 4.2K InSb bolometer (QMC Instruments Ltd.)

The FIR method was used in HMFL Nijmegen and the CW method in NHMFL Tallahassee.
Radiation propagated either parallel or perpendicular to the applied magnetic field in Faraday or Voigt configurations, respectively.

THz radiation was guided to the sample from the top of  the liquid helium cryostat and from the sample to the detector with light pipes.
In the FIR setup, the bolometer was placed in the tail of the sample cryostat below the center of the magnet.
In the CW setup, the bolometer was placed in a separate cryostat a few meters away from the magnet.
Radiation was either unpolarized or polarized by a wire grid located a few millimeters from the sample surface in the incident THz beam.
Sample temperature was maintained between 2 and 8\,K.

In the Faraday configuration, each sample was measured in a magnetic field perpendicular to the cutting plane. 
In the Voigt configuration, each sample was measured in  magnetic fields along two principal directions in the cutting plane. 
After the sample was cooled in zero field, measurements were carried out in different magnetic fields or by sweeping the field, both with a fixed field orientation.

\subsection{Spectra}
\subsubsection{FIR spectroscopy}
\begin{figure}
	\includegraphics[width=\columnwidth]{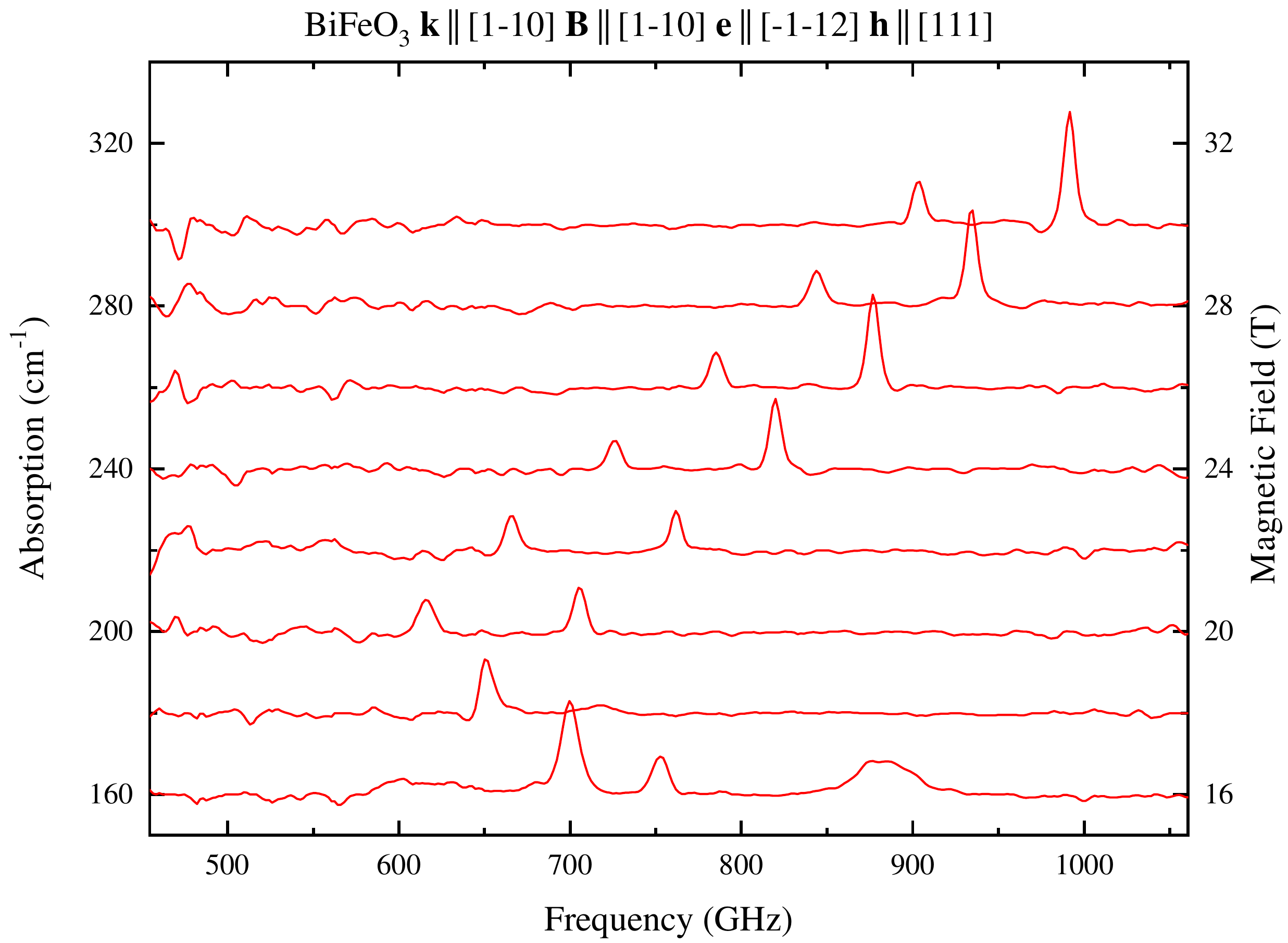} 
	\caption{THz absorption spectra $\alpha_B(\nu)$ for the magnetic field direction $\bdc/ \parallel [1,-1,0]$. 
	The vertical shift of the spectra is proportional to $B$.}
	\label{fig:BFO_(1-10)_B1-10_e-1-12_b111}
\end{figure}

\begin{figure}
	\includegraphics[width=\columnwidth]{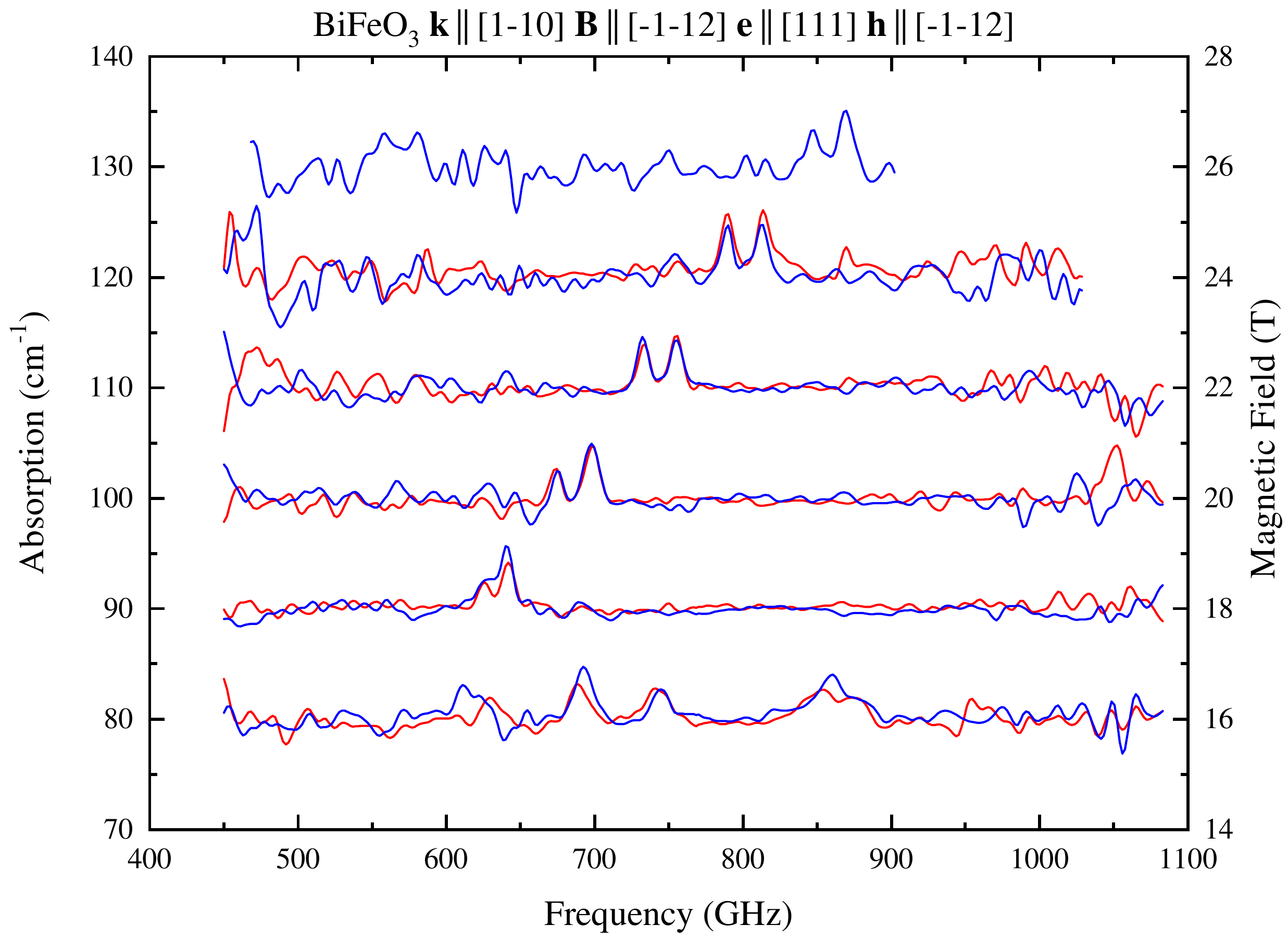} 
	\caption{THz absorption spectra $\alpha_B(\nu)$ for the magnetic field direction $\bdc/ \parallel [-1,-1,2]$. 
	The vertical shift of the spectra is proportional to $B$.
	Red and blue lines show the spectra measured in positive and negative magnetic fields.
	}
	\label{fig:BFO_(1-10)_B-1-12_e111_b-1-12}
\end{figure}

\begin{figure}
	\includegraphics[width=\columnwidth]{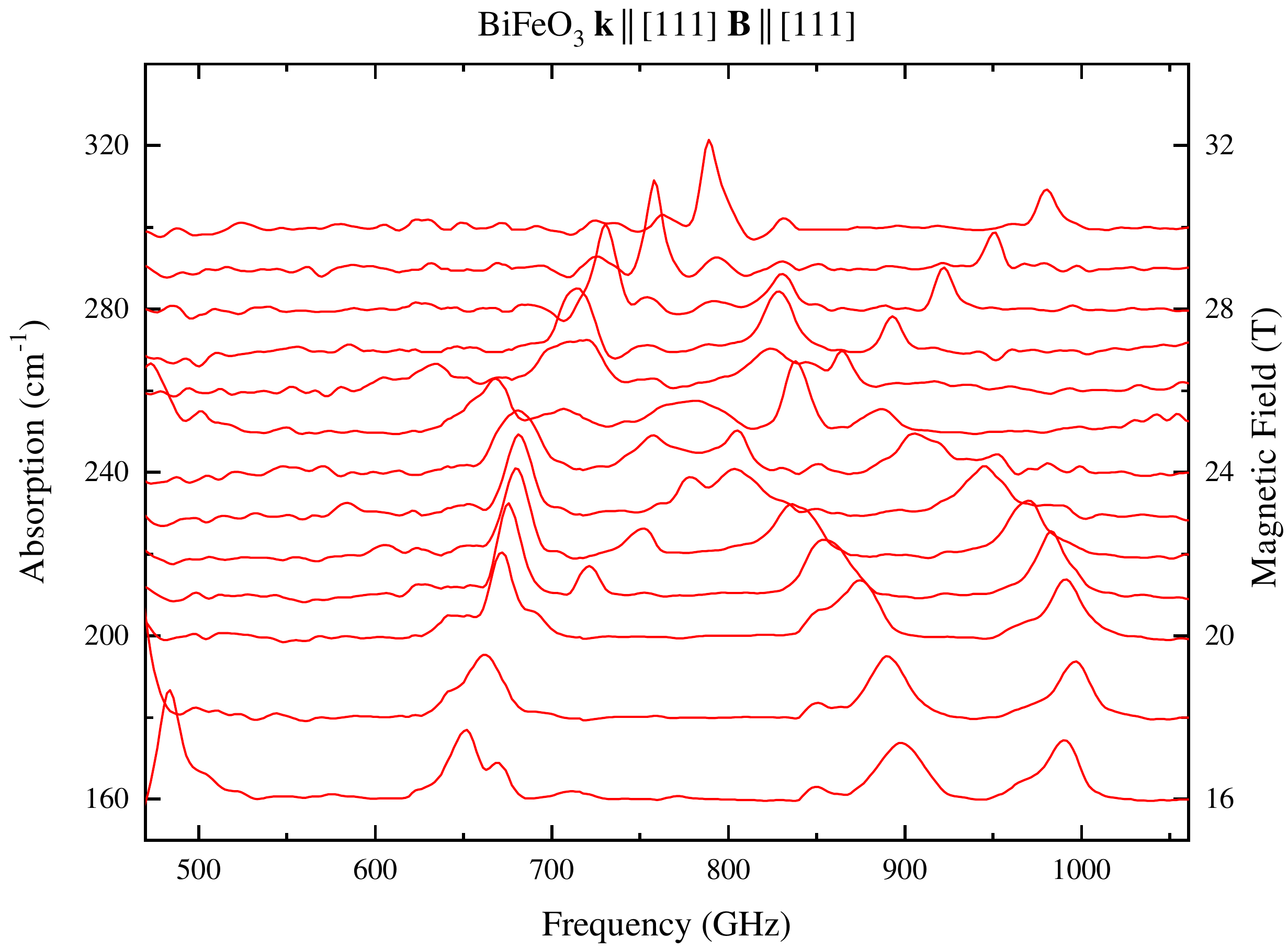} 
	\caption{THz absorption spectra $\alpha_B(\nu)$ for the magnetic field direction $-\vect{k}\parallel\bdc/ \parallel [1,1,1]$ measured with unpolarized radiation. 
	The vertical shift of the spectra is proportional to $B$.
	}
	\label{fig:BiFeO3_B111_Nijmegen_abs}
\end{figure}

The absorption spectra $\alpha_B(\nu)$ obtained by FIR spectroscopy were calculated from the difference \cite{SWBook2018} $\alpha_B(\nu)-\alpha_{B=0}(\nu) =-\ln[I_B( \nu)/I_{B=0}(\nu)]/d$, where $I_B(\nu)$ is the transmitted intensity of radiation for sample thickness $d$.
Mode frequencies were determined by fitting the spectra with a Gaussian line shape.
The FIR spectra are shown in Fig.\,\ref{fig:BFO_(1-10)_B1-10_e-1-12_b111}, \ref{fig:BFO_(1-10)_B-1-12_e111_b-1-12} and \ref{fig:BiFeO3_B111_Nijmegen_abs}.
The radiation polarization, i.e. the directions of the THz electric $\mathbf{e}$ and magnetic $\mathbf{h}$ fields, are given in the figure titles.

\subsubsection{CW spectroscopy}

The transmission spectra $t_\nu(B)$ measured by CW spectroscopy were calculated using $t_\nu(B) =I_\nu(B)/\bar{I}_\nu$, where $\bar{I}_\nu$ is the mean value of $I_\nu(B)$ over the field sweep, typically from 0 to 35\,T.
Mode frequencies were determined from the transmission line minima. 

Due to the distortion of the transmission line shape in the CW method, the transmission minimum may not correspond to the magnetic field with the strongest absorption.
Because the $v_1$ data was taken only with the CW method, its lineshape is more distorted and the data points are more scattered than for the $v_2$ data in Fig.\,\ref{fig:fit_R3c} 
In addition, the magnetic-field dependence of $v_1$ is less steep than that of $v_2$.  Hence, its line position is less accurate in the magnetic-field scan.

\begin{figure}
	\includegraphics[width=\columnwidth]{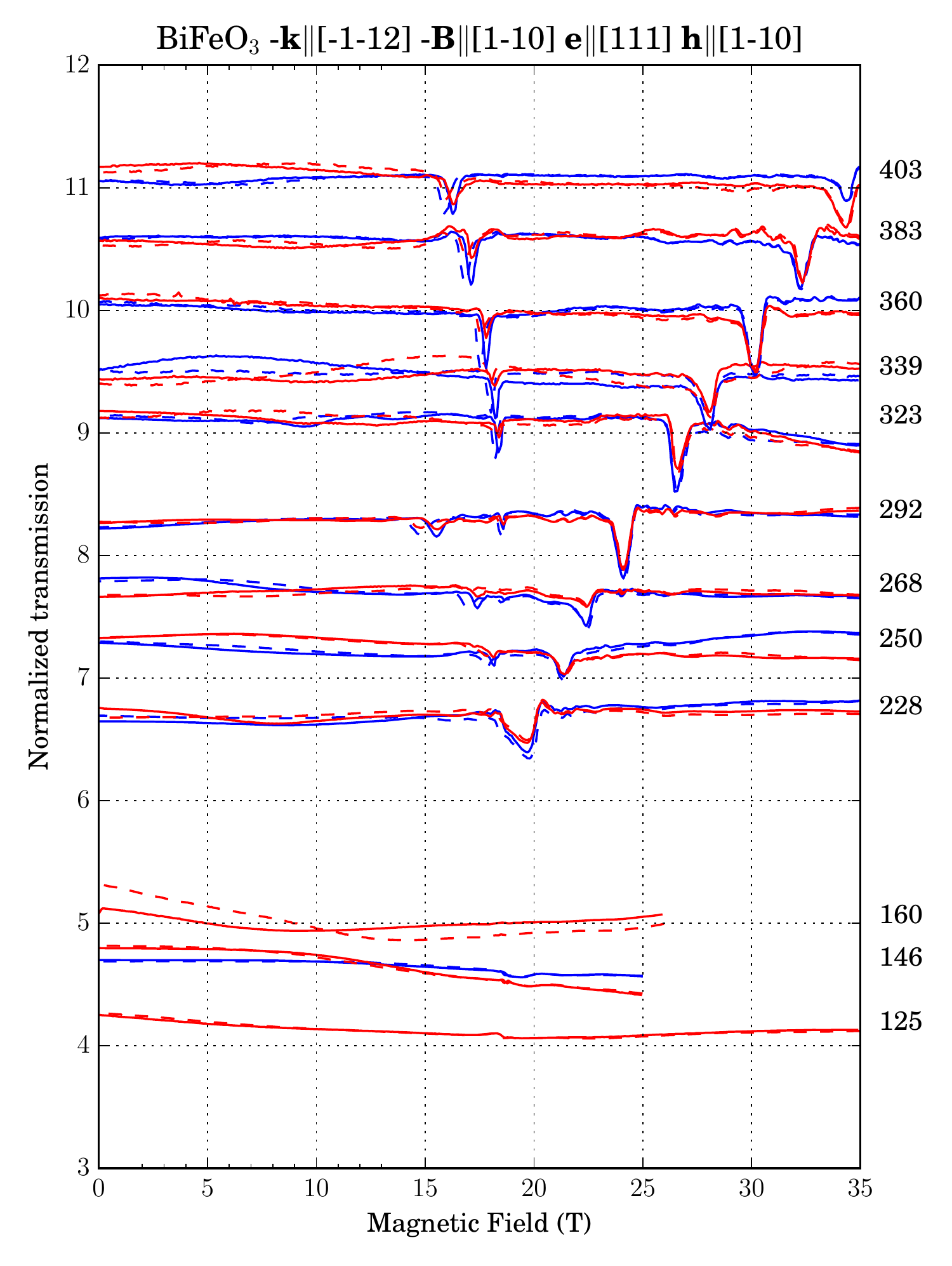} 
	\caption{\label{fig:bfo_-1-12_b-110_e111}
	}
\end{figure}

\begin{figure}
	\includegraphics[width=\columnwidth]{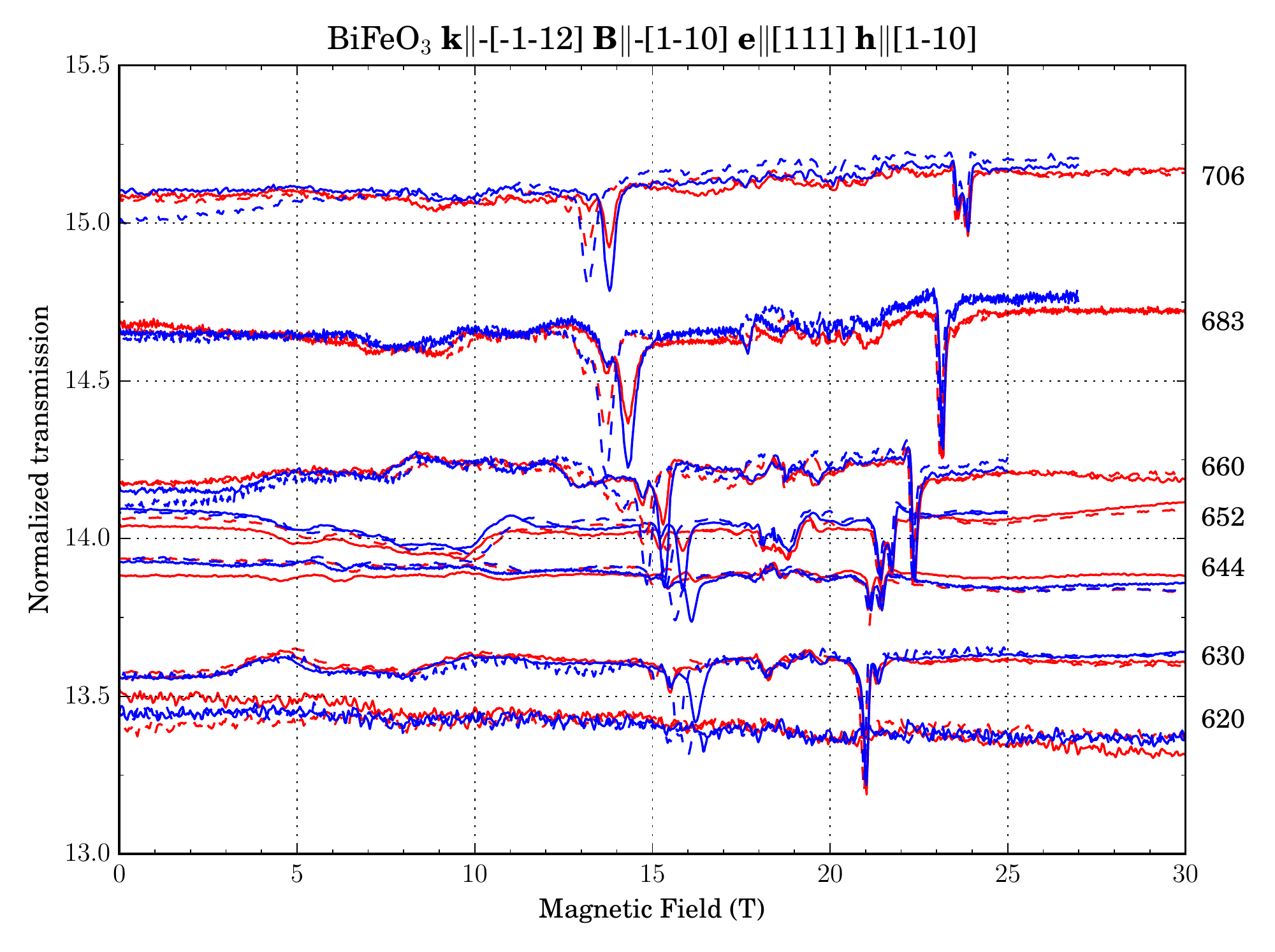} 
	\caption{}
	\label{fig:bfo_-1-12_-b1-10_e111}
\end{figure}

\begin{figure}
	\includegraphics[width=\columnwidth]{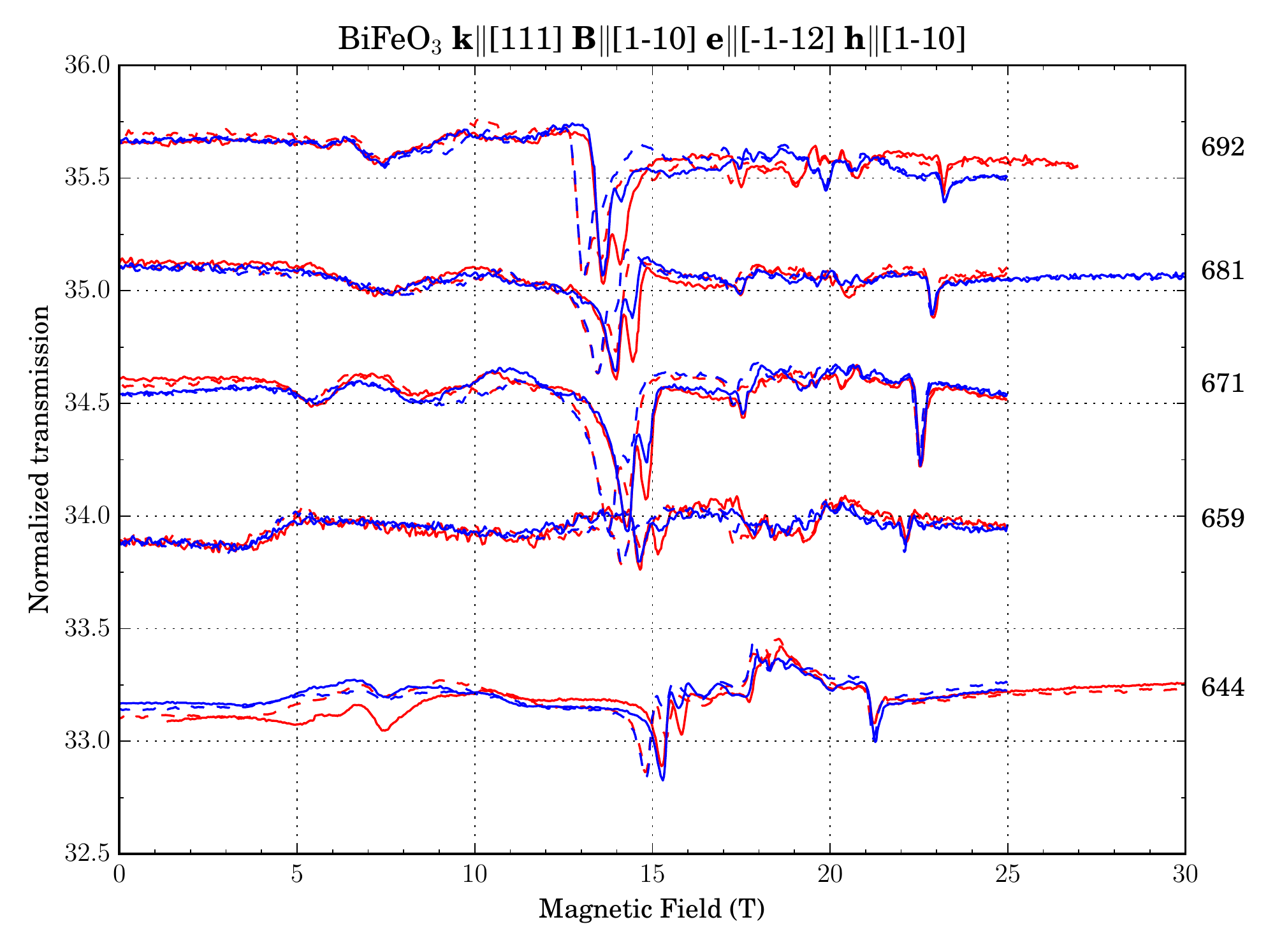} 
	\caption{\label{fig:bfo_111_b1-10_e-1-12}}
\end{figure}

\begin{figure}
	\includegraphics[width=\columnwidth]{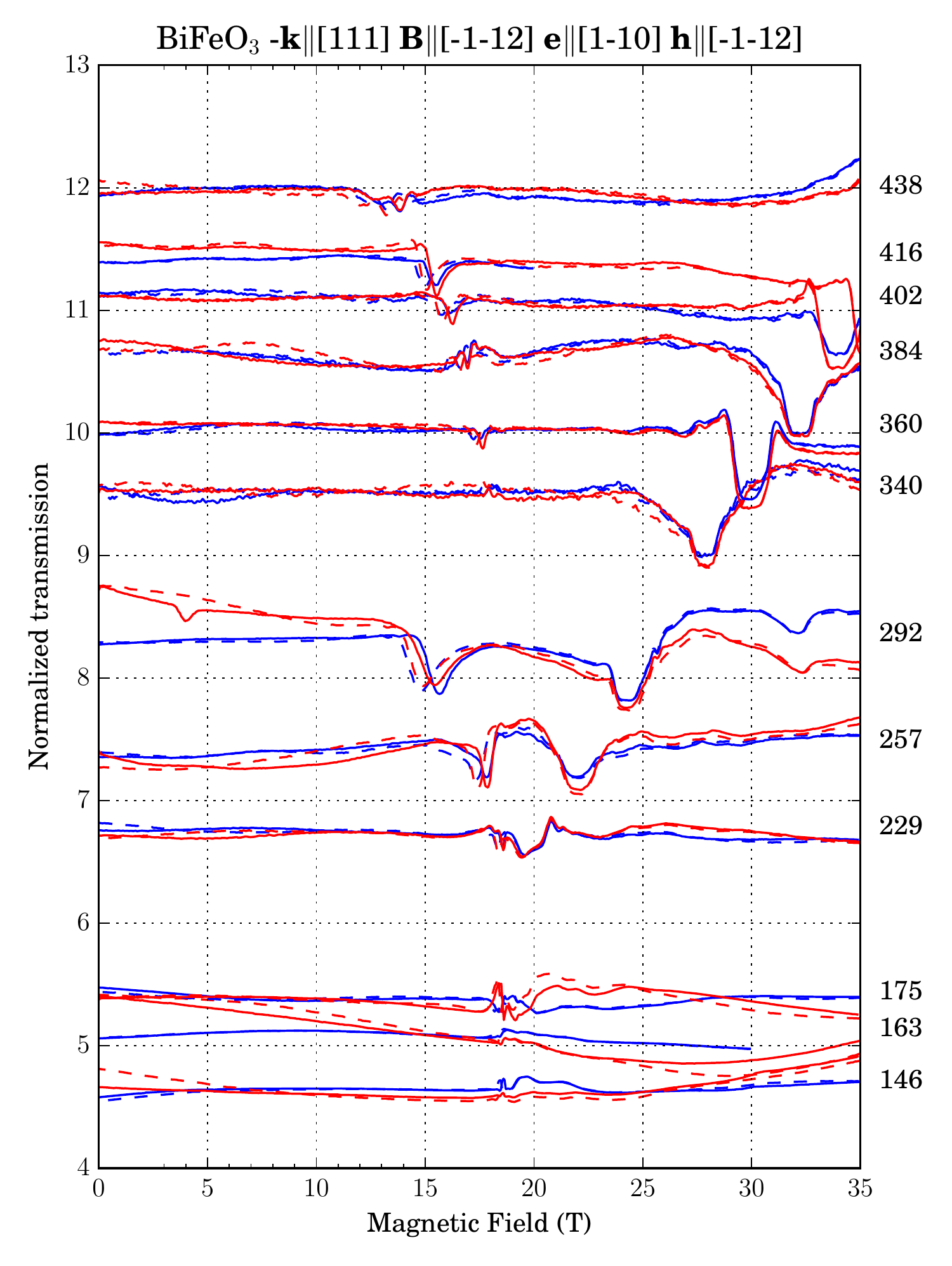}
	\caption{\label{fig:bfo_111_b-1-12_e1-10}}
\end{figure}

\begin{figure}
	\includegraphics[width=\columnwidth]{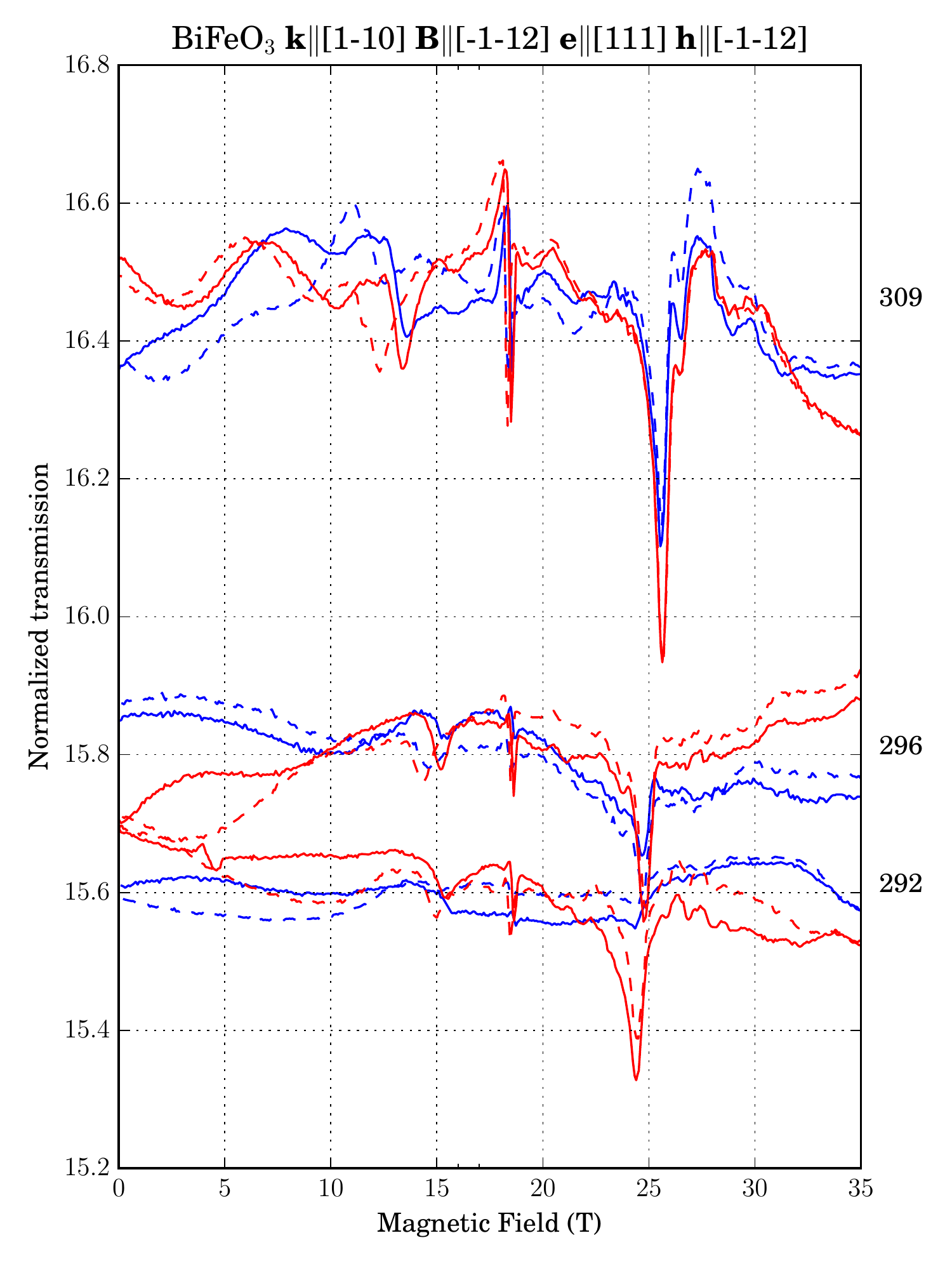} 
	\caption{\label{fig:bfo_1-10_b-1-12_e111}} 
\end{figure}

\begin{figure}
	\includegraphics[width=\columnwidth]{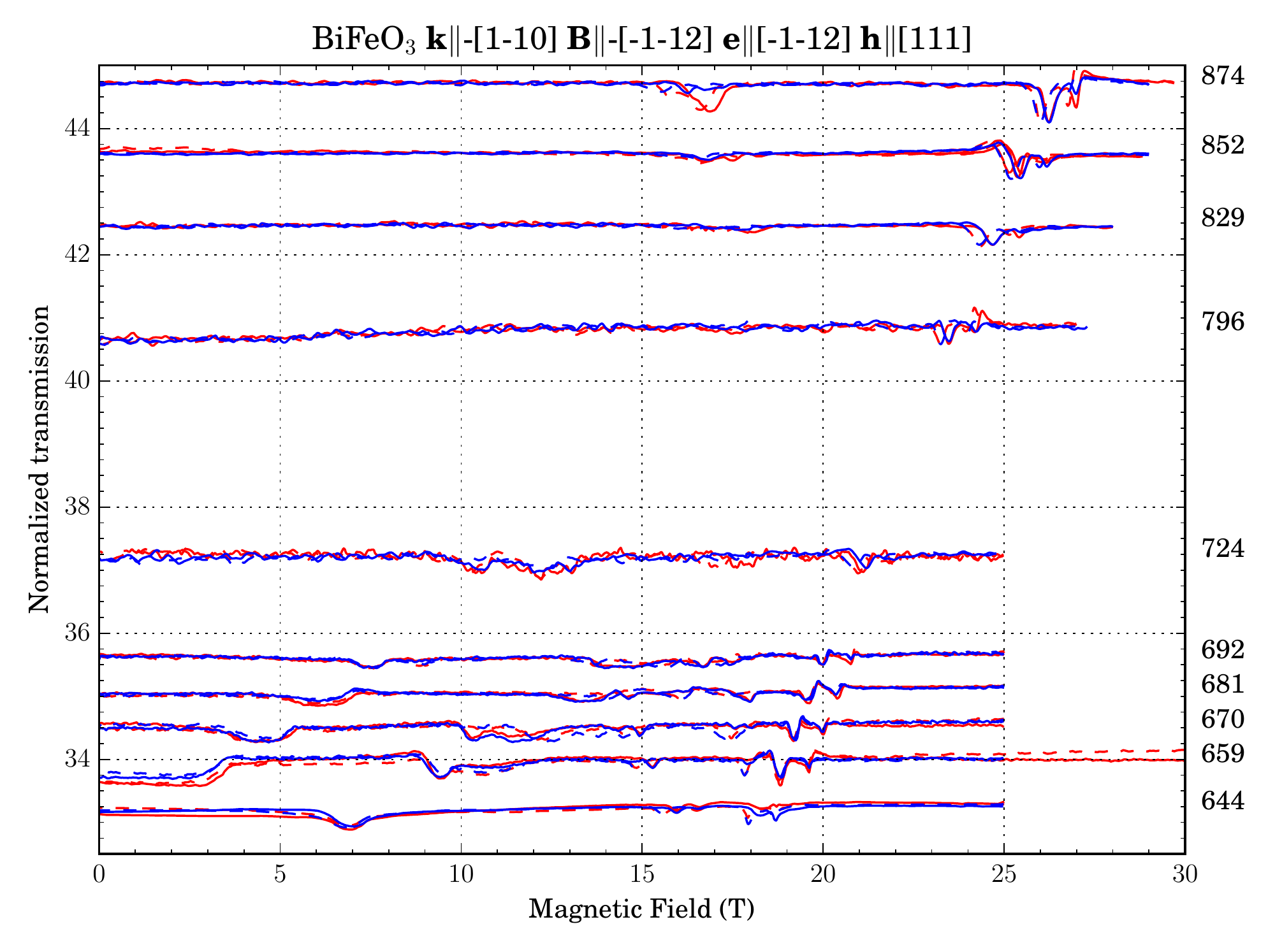}
	\caption{\label{fig:bfo_1-10_b11-2_e-1-12}}
\end{figure}

\begin{figure}
	\includegraphics[width=\columnwidth]{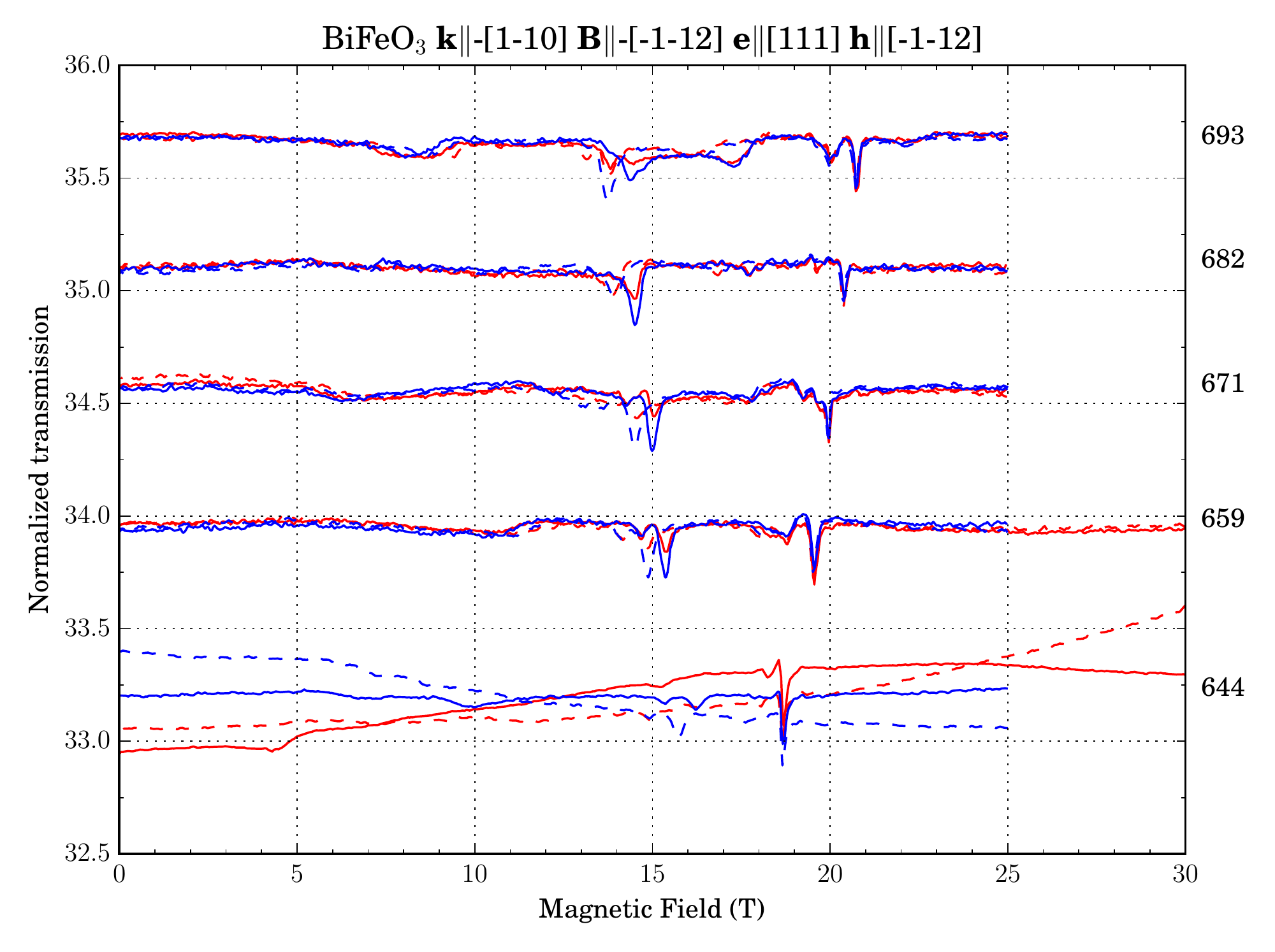} 
	\caption{\label{fig:bfo_1-10_b11-2_e111}}
\end{figure}

\begin{figure}
	\includegraphics[width=\columnwidth]{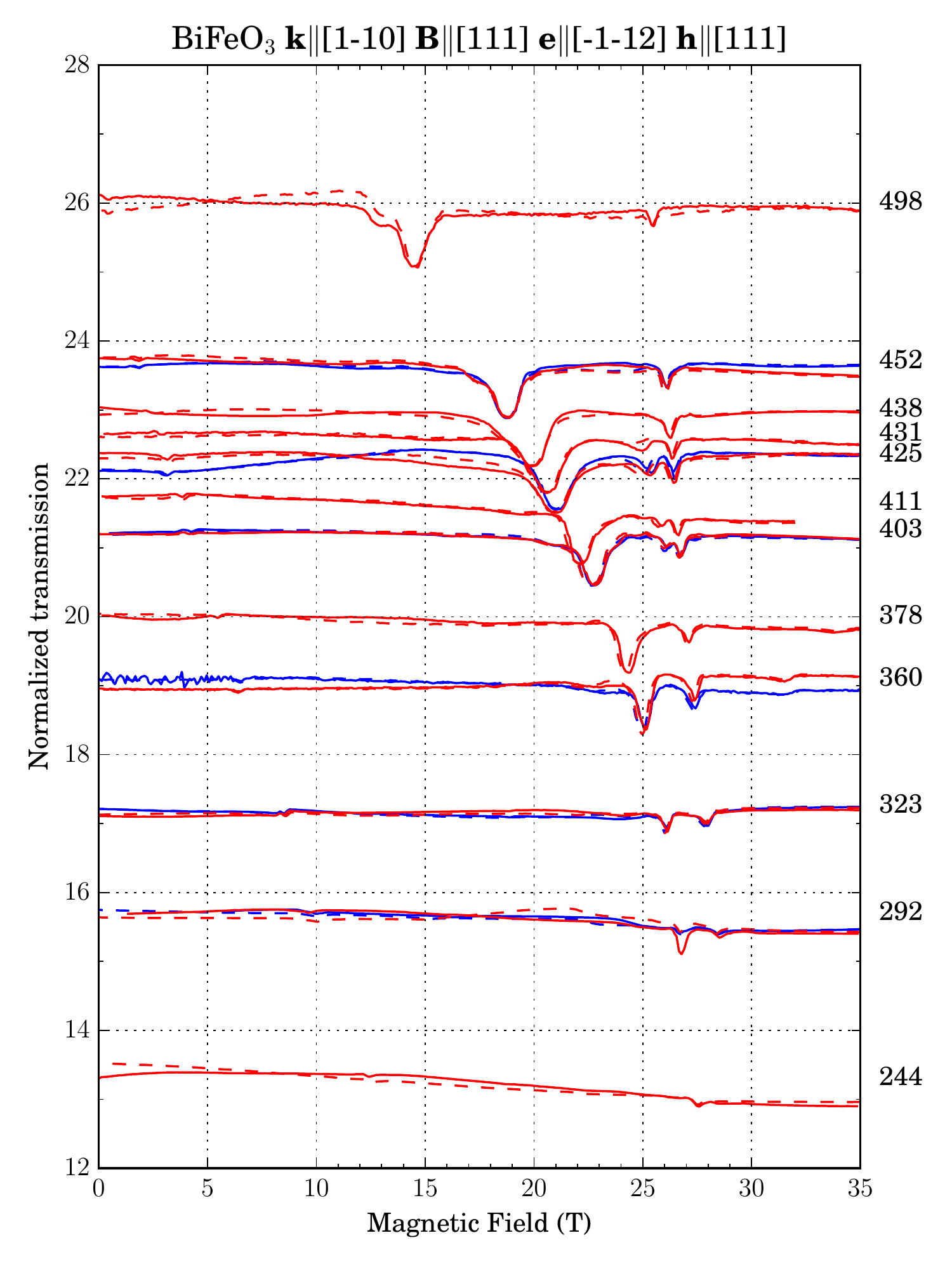} 
	\caption{\label{fig:bfo_1-10_b111_e-1-12}}
\end{figure}

\begin{figure}
	\includegraphics[width=\columnwidth]{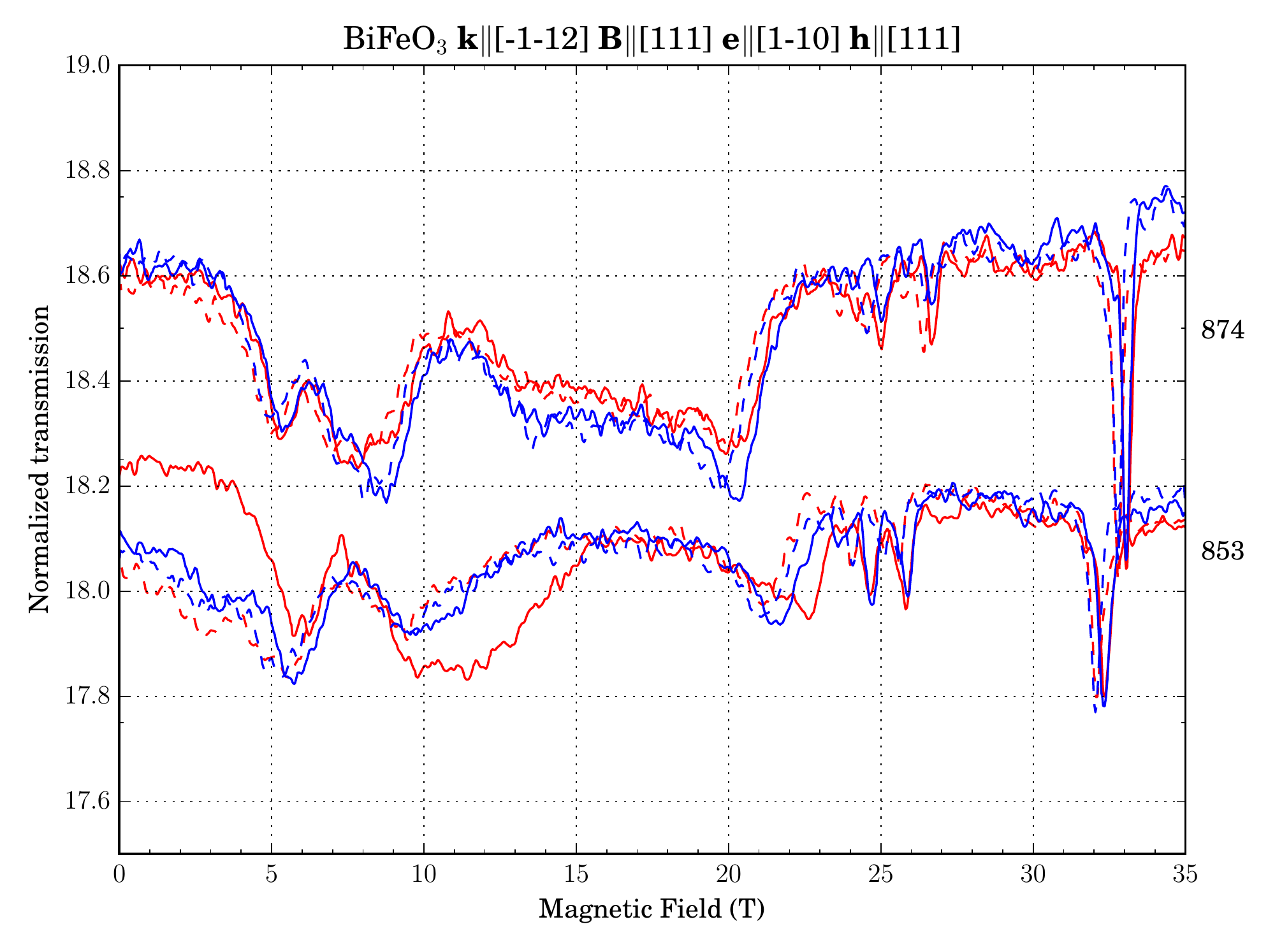}
	\caption{\label{fig:bfo_-1-12_b111_e1-10}}
\end{figure}

\begin{figure}
	\includegraphics[width=\columnwidth]{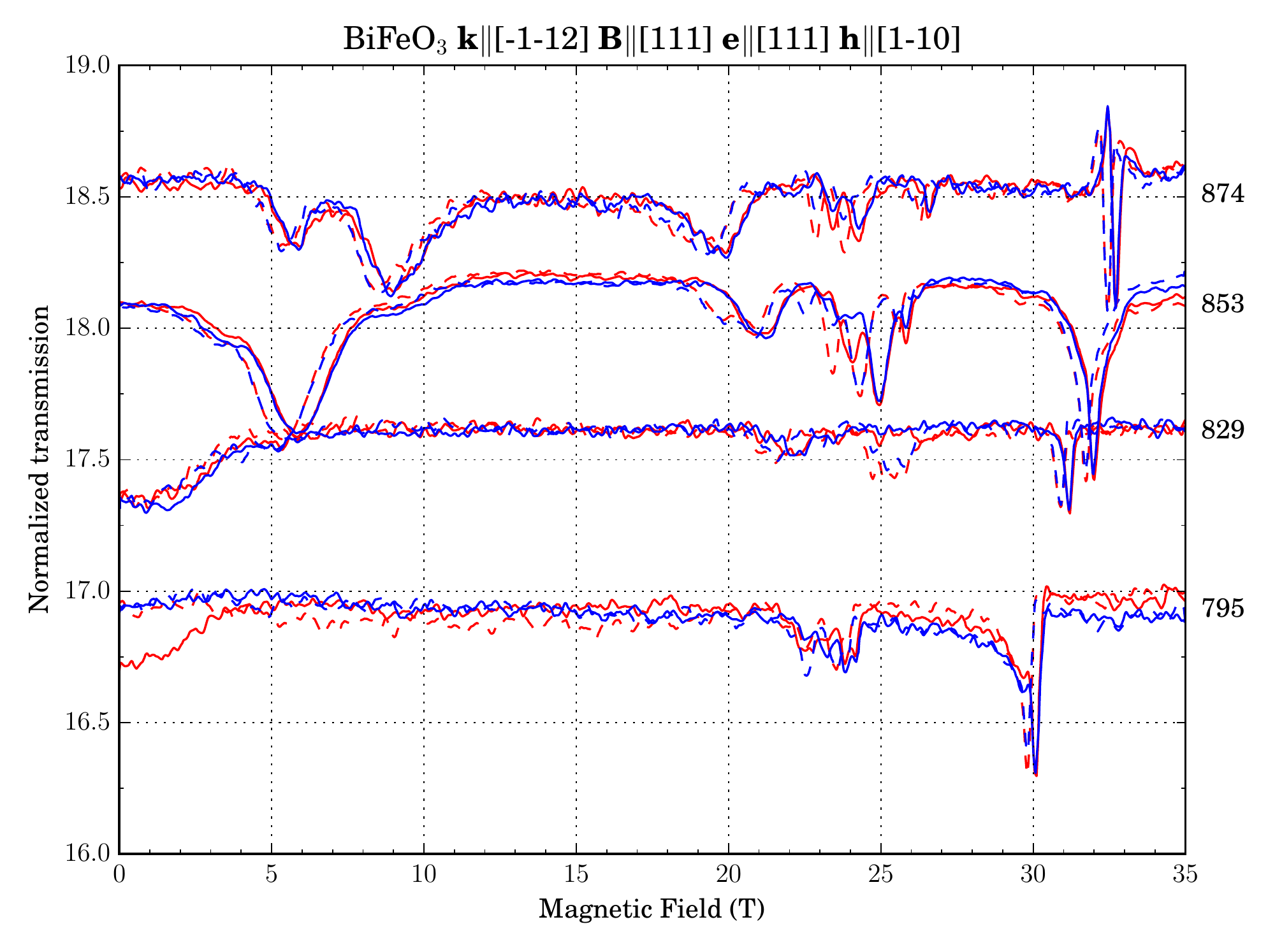} 
	\caption{\label{fig:bfo_-1-12_b111_e111-tallahassee}}
\end{figure}

\begin{figure}
	\includegraphics[width=\columnwidth]{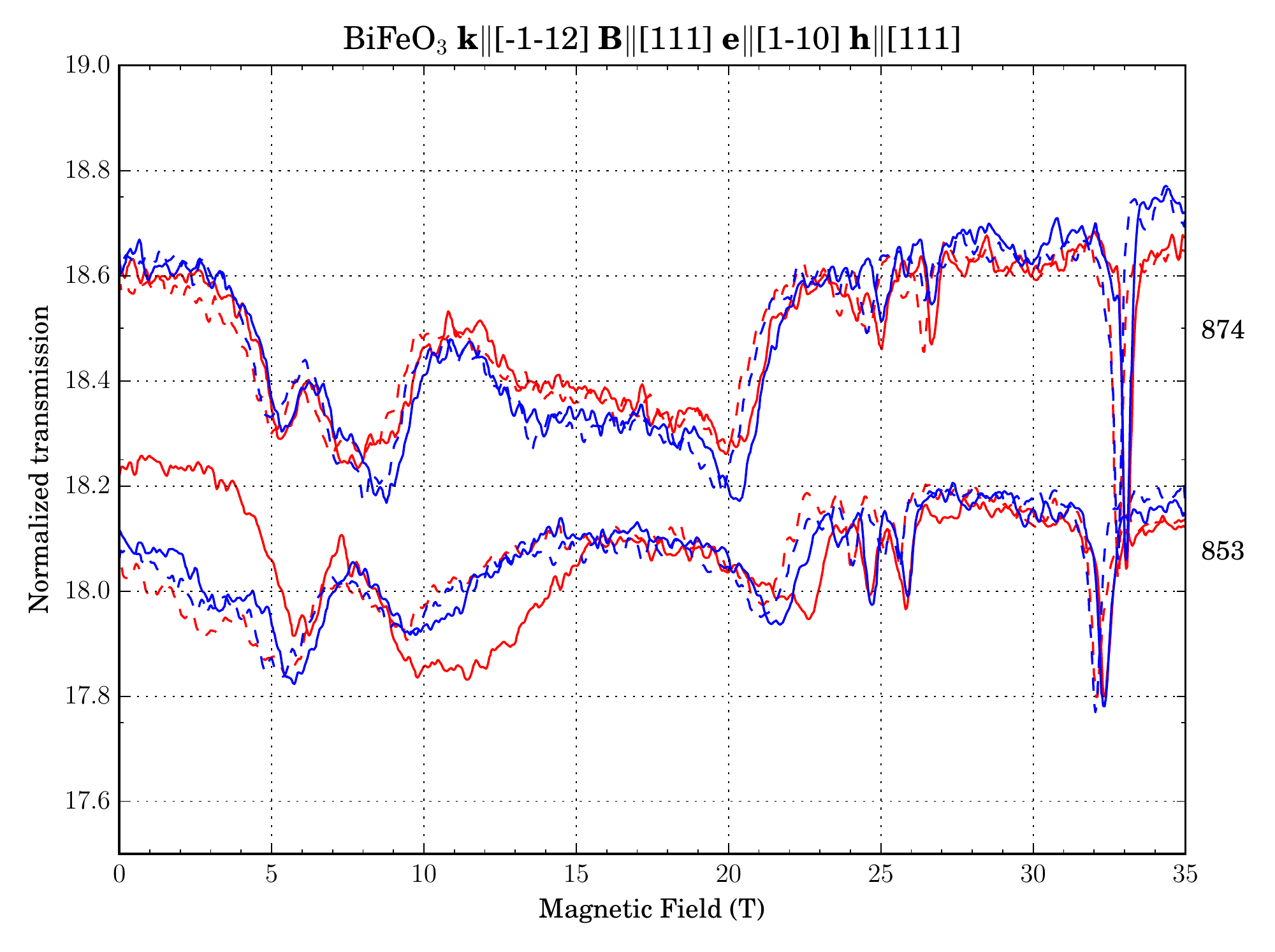} 
	\caption{\label{fig:bfo_-1-12_b111_e1-10_v2}}
\end{figure}

The titles of Figs.\,\ref{fig:bfo_-1-12_b-110_e111}-\ref{fig:bfo_-1-12_b111_e1-10_v2} show the direction $\vect{k}$ of light propagation, the direction of the applied magnetic field \bdc/, and the directions of the radiation electric ($\vect{e}$) and magnetic ($\vect{h})$ field components.
The red (blue) line is for $ B >0$ ($B<0$).
The solid line is for increasing fields, $\mathrm{d}|B|/\mathrm{d} t>0$, and the dashed line for decreasing fields, $\mathrm{d}|B|/\mathrm{d} t<0$.
In case of hysteresis, the average of the up and down sweep transmission line minima determined the mode resonance field.  
The frequency in GHz units is given on the right side of each figure.

\vfill\eject


\hskip .3cm


\bibliographystyle{apsrev}

\end{document}